\documentclass[a4paper,12pt]{article}
\usepackage[margin=2.5cm]{geometry}
\usepackage{graphicx}
\usepackage{setspace}
\usepackage{caption}
\usepackage{subcaption}
\usepackage{indentfirst}
\usepackage{footmisc}
\usepackage{amsmath}
\usepackage{amssymb}
\usepackage{enumerate}
\usepackage[nosort]{cite}
\usepackage[svgnames]{xcolor}
\usepackage[colorlinks=true,
            allcolors=.,
            bookmarksnumbered=true,
            pdfpagemode=UseNone,
            pdfstartview=FitH]{hyperref}

\newcommand{\dif}{\text{d}}
\DeclareMathOperator{\am}{am}
\DeclareMathOperator{\sn}{sn}
\DeclareMathOperator{\sd}{sd}

\numberwithin{equation}{section}

\makeatletter
\renewcommand\section{\@startsection {section}{1}{\z@}%
	{-3.5ex \@plus -1ex \@minus -.2ex}%
	{2.3ex \@plus.2ex}%
	{\normalfont\sffamily\bfseries}}
\renewcommand\subsection{\@startsection{subsection}{2}{\z@}%
	{-3.25ex\@plus -1ex \@minus -.2ex}%
	{1.5ex \@plus .2ex}%
	{\normalfont\sffamily\slshape}}
\makeatother

\begin{document}

\thispagestyle{empty}
\vbox{}
\vspace{2cm}

\begin{center}
  {\sffamily\LARGE{Spherical orbits around a Kerr black hole
  }}\\[16mm]
  {\sffamily Edward Teo}
  \\[6mm]
    {\sffamily\slshape\selectfont
      Department of Physics, National University of Singapore, Singapore
    }\\[15mm]
\end{center}
\vspace{2cm}
	
\centerline{\sffamily\bfseries Abstract}
\bigskip
\noindent  A special class of orbits known to exist around a Kerr black hole are spherical orbits---orbits with constant coordinate radii that are not necessarily confined to the equatorial plane. Spherical time-like orbits were first studied by Wilkins almost 50 years ago. In the present paper, we perform a systematic and thorough study of these orbits, encompassing and extending previous works on them. We first present simplified forms for the parameters of these orbits. The parameter space of these orbits is then analysed in detail; in particular, we delineate the boundaries between stable and unstable orbits, bound and unbound orbits, and prograde and retrograde orbits. Finally, we provide analytic solutions of the geodesic equations, and illustrate a few representative examples of these orbits.

\newpage

\section{Introduction}

Of all the known exact solutions of Einstein's equations, the Kerr solution \cite{Kerr:1963} describing a rotating black hole remains the most important one from an astrophysical viewpoint. The recent spectacular detection of gravitational waves from the merger of a pair of black holes \cite{Abbott:2016}, and the direct imaging of a supermassive black hole \cite{Akiyama:2019}, have elevated the study of the Kerr black hole to an observational science. This has given new impetus to the study of time-like and null geodesics around a Kerr black hole (see, e.g., \cite{Rana:2019,Kapec:2019,Gralla:2019,Stein:2019,Compere:2020,Rana:2020}).

The study of geodesics around a Kerr black hole has a long history. It essentially started in 1968, with Carter's remarkable discovery of a Killing tensor for the Kerr space-time \cite{Carter:1968}. This has had two important (related) consequences: The first is the existence of a new conserved quantity associated to each geodesic, the so-called Carter's constant. The second is that the geodesic equations can be written as a set of first-order differential equations.

The existence of Carter's constant means that geodesics around a Kerr black hole are characterised by a total of three (non-trivial) conserved quantities. The other two are just the energy and angular momentum of the particle. The physical interpretation of Carter's constant is less obvious; it turns out that it governs the motion of geodesics in the polar direction. If it is zero, geodesics initially moving in the equatorial plane will remain in the equatorial plane. In general, however, non-equatorial motion is allowed. Depending on the sign of Carter's constant, different orbital behaviour can result. A recent classification of the different types of orbital behaviour allowed can be found in \cite{Compere:2020}.

The fact that the geodesic equations can be written as a set of first-order differential equations means that they can be readily solved. Traditionally, a numerical procedure, such as the fourth-order Runge-Kutta method, was used to solve them. However, in 2003, Mino showed how two of these equations can be decoupled through the introduction of a new time parameter \cite{Mino:2003}. This then allows the geodesic equations to be analytically solved in terms of this parameter using, for example, elliptic integrals in the Legendre canonical form \cite{Fujita:2009}. A review of the known analytic solutions of geodesics in the Kerr and other related space-times can be found in \cite{Hackmann:2015}.

Amongst the different types of possible non-equatorial orbits, those with constant coordinate radii are distinguished, just as circular orbits are distinguished in the class of equatorial orbits. Such orbits are known as {\it spherical orbits\/}. This special class of geodesics is obviously simpler to analyse than more general ones. Yet spherical orbits remain astrophysically relevant. For example, they mark the threshold between non-equatorial orbits that plunge into the black hole, and those that do not. Such threshold orbits play an important role in modelling the capture of matter and light by the black hole.

Spherical time-like orbits around a Kerr black hole were first studied by Wilkins \cite{Wilkins:1972}, almost 50 years ago. In his groundbreaking work, he analysed many of their properties. In particular, he plotted out the parameter space of stable spherical orbits around an extremal Kerr black hole. These spherical orbits have been further studied by various authors over the years (see, e.g., \cite{Johnston:1974,Stoghianidis:1987,Hughes:1999,Hughes:2001,Kraniotis:2004,Fayos:2007,Hackmann:2010,Grossman:2011,Hod:2013,Rana:2019,Stein:2019,Compere:2020}).  

As it turns out, several different parameterisations of the orbital parameters have appeared in the literature. Due to the nature of the equations being solved, they have tended to take rather complicated algebraic forms. This has in turn obscured the full picture of the parameter space of these spherical orbits. One of the aims of this paper is to present simplified forms of the orbital parameters. This will allow us to find the appropriate ranges for the parameters, and come up with an understanding of the whole parameter space. In particular, we will be able to extend Wilkins' parameter space to include unstable orbits, both bound and unbound.

For completeness, we will then provide analytic solutions of the geodesic equations in terms of the Mino parameter \cite{Mino:2003}. This will allow for the efficient plotting of the spherical orbits, once their parameters have been chosen. Ultimately, it is hoped that the results in this paper---together with the author's earlier work on spherical {\it photon\/} orbits \cite{Teo:2003}---would be useful for readers interested in all types of spherical orbits around a Kerr black hole.

This paper is organised as follows: We begin in Sec.~\ref{section 2} with a brief review of the relevant geodesic equations, in particular focussing on the one governing motion in the polar direction. In Sec.~\ref{section 3}, the conditions for a spherical time-like orbit are solved, and the energy and angular momentum of the orbit are expressed in terms of its radius and its value of Carter's constant. The appropriate ranges for the latter two parameters are found. In Sec.~\ref{section 4}, the properties of these spherical orbits are analysed. In particular, we find analytic expressions for the boundaries of the parameter space separating stable and unstable orbits, bound and unbound orbits, and prograde and retrograde orbits. In Sec.~\ref{section 5}, we provide analytic solutions of the geodesic equations in terms of the Mino parameter using elliptic integrals and Jacobi elliptic functions.  We then illustrate a few representative examples of spherical orbits in Sec.~\ref{section 6}. The paper ends off with two appendices. In the first appendix, we consider the special class of so-called horizon-skimming orbits \cite{Wilkins:1972}, which appear to lie on the event horizon of the extremal Kerr black hole. In the second appendix, we provide analytic solutions of the geodesic equations for spherical photon orbits, thereby supplementing the results of \cite{Teo:2003} in which these equations were solved numerically.

\section{Geodesic equations}
\label{section 2}

In standard Boyer--Lindquist coordinates $(t,r,\theta,\phi)$, the Kerr black hole has the line element
\begin{align}
\label{Kerr}
\dif s^2&=-\left(1-\frac{2Mr}{\Sigma}\right)\dif t^2-\frac{4Mr}{\Sigma}\,a\sin^2\theta\,\dif t\dif\phi+\Sigma\left(\frac{\dif r^2}{\Delta}+\dif\theta^2\right)\cr
&\qquad+\left(r^2+a^2+\frac{2Mr}{\Sigma}\,a^2\sin^2\theta\right)\sin^2\theta\,\dif\phi^2,
\end{align}
where
\begin{subequations}
\begin{align}
\Sigma&\equiv r^2+a^2\cos^2\theta\,,\\
\Delta&\equiv r^2-2Mr+a^2.
\end{align}
\end{subequations}
The parameters $M$ and $a$ are the black hole's mass and angular momentum per unit mass, respectively, and are assumed to lie in the range $0<a\leq M$. The event horizon of the black hole is located at the radius $r=r_\text{H}$, where
\begin{align}
r_\text{H}\equiv M+\sqrt{M^2-a^2}
\end{align}
is the larger root of $\Delta$. Since we are only interested in particle motion outside the event horizon, $r$ is assumed to lie in the range $r_\text{H}<r<\infty$. The other coordinates take the usual ranges.

In place of the coordinate $\theta$, it turns out to be useful to define the coordinate $u\equiv\cos\theta$, with $-1\leq u\leq1$. The four geodesic equations governing the motion of a particle in the space-time (\ref{Kerr}) are then \cite{Carter:1968,Wilkins:1972}
\begin{subequations}
\label{eom}
\begin{align}
\label{r_eom}
\Sigma\frac{\dif r}{\dif\tau}&=\pm\sqrt{R(r)}\,,\\
\label{u_eom}
\Sigma\frac{\dif u}{\dif\tau}&=\pm\sqrt{V(u)}\,,\\
\label{phi_eom}
\Sigma\frac{\dif\phi}{\dif\tau}&=\frac{\Phi}{1-u^2}+\frac{a}{\Delta}(2MrE-a\Phi)\,,\\
\label{t_eom}
\Sigma\frac{\dif t}{\dif\tau}&=-a^2E(1-u^2)+\frac{1}{\Delta}\big[E(r^2+a^2)^2-2Mra\Phi\big]\,,
\end{align}
\end{subequations}
where 
\begin{subequations}
\label{RU}
\begin{align}
\label{R}
R(r)&\equiv\big(E^2-\mu^2\big)r^4+2M\mu^2r^3+\big[a^2(E^2-\mu^2)-Q-\Phi^2\big]r^2\cr
&\qquad+2M\big[(aE-\Phi)^2+Q\big]r-a^2Q\,,\\
V(u)&\equiv a^2\big(\mu^2-E^2\big)u^4-\big[a^2(\mu^2-E^2)+Q+\Phi^2\big]u^2+Q\,.
\end{align}
\end{subequations}
They are a set of first-order differential equations with respect to an affine parameter $\tau$ along the geodesic. There are three constants of motion appearing in these equations: $E$ and $\Phi$ are the particle's energy and angular momentum about the $\phi$-axis, respectively, while $Q$ is Carter's constant. There is also a trivial fourth constant of motion $\mu$, which is the rest mass of the particle. Since the focus of this paper is on time-like particles, we may take $\mu=1$ without loss of generality. 

Note that the right-hand sides of the geodesic equations (\ref{r_eom}) and (\ref{u_eom}) are square roots. The requirement that they are real will be a first step towards relating the constants of motion to the behaviour of the orbits, as we shall briefly review in the following two subsections \cite{Carter:1968,Wilkins:1972}.

\subsection{Geodesic equation for \texorpdfstring{$r$}{r}}
\label{r geodesic}

From the geodesic equation for $r$, (\ref{r_eom}), we see that the physically allowed ranges for $r$ can only occur when $R(r)$ is non-negative. If $r$ is allowed in a finite range outside the event horizon, then the corresponding orbit is said to be {\it bound\/}; if $r$ is allowed in a semi-infinite range outside the event horizon (so that it extends to infinity), then the orbit is said to be {\it unbound\/}. The (finite) boundaries of these ranges are given by the roots of $R(r)$, which is a quartic equation in $r$. In \cite{Wilkins:1972}, Wilkins used Descartes' rule of signs, which links the number of positive roots of a polynomial to the number of sign changes of its coefficients, to deduce the number of roots of $R(r)$ lying outside the event horizon. In particular, he showed that bound orbits can only occur if $E^2<1$. When $E^2\geq1$, only unbound orbits are allowed. Orbits for which $E^2=1$ are also known as {\it marginally bound\/} orbits \cite{Bardeen:1972}.

We remark that constant-radii orbits---namely circular or spherical orbits---may either be bound or unbound. In the unbound case, a constant-radius orbit is necessarily unstable, and an outward perturbation will cause it to escape to infinity. In the bound case, a constant-radius orbit may either be stable or unstable. If it is unstable, a radial perturbation will turn it into an eccentric orbit, whose radius varies between two finite values.

\subsection{Geodesic equation for \texorpdfstring{$u$}{u}}
\label{u geodesic}

Similarly, from the geodesic equation for $u$, (\ref{u_eom}), we see that the physically allowed ranges for $u$ occur when $V(u)$ is non-negative. Since $V(u\mathop{=}\pm1)=-\Phi^2\leq0$, the orbits can only reach the poles $|u|=1$ if $\Phi=0$. In general, we require the existence of at least one part of the range $[-1,1]$ for which $V(u)\geq0$. The boundaries of this physically allowed range are given by the roots of $V(u)$, which is biquadratic in $u$, or quadratic in $w\equiv u^2$. Its roots, in terms of the new variable $w$, are\footnote{Here, and subsequently, the first subscript refers to the upper sign, while the second subscript refers to the lower sign.}
\begin{align}
\label{w12}
w_{1,2}^{\phantom{}}\equiv u_{1,2}^2&=\frac{1}{2a^2(1-E^2)}\bigg[a^2(1-E^2)+Q+\Phi^2\cr
&\qquad\mp\sqrt{\big(a^2(1-E^2)+Q+\Phi^2\big)^2-4a^2Q(1-E^2)}\bigg]\,.
\end{align}
The ranges of these two roots depend in particular on the sign of $Q$, as well as whether $E^2<1$ or $E^2>1$.\footnote{The marginally bound case $E^2=1$ will not be treated separately, as it can be obtained by taking the limit $E^2\rightarrow1^-$ of the bound case. Expressions for $w_{1,2}$ in this limit can be found in Eq.~(\ref{E=1_limit}) below.} We have the following cases \cite{Carter:1968}:

\begin{enumerate}[(i)]
\item{\underline{$Q>0$}:}\enskip Since $V(w\mathop{=}0)=Q$, and recalling that $V(w\mathop{=}1)\leq0$, it follows that exactly one of the two roots $w_{1,2}$ will lie in the range $(0,1]$. Indeed, it can be checked that $0<w_1\leq1<w_2$ when $E^2<1$, and $w_2<0<w_1\leq1$ when $E^2>1$. In either subcase, the physically allowed range for $u$ is given by $|u|\leq u_1=\sqrt{w_1}$. It describes an orbit that crosses the equatorial plane $u=0$ repeatedly, oscillating between the latitudes $\pm u_1$. 

\item{\underline{$Q<0$}:}\enskip If $E^2<1$, it follows from the general shape of a quadratic function with positive leading coefficient that both roots $w_{1,2}$ must lie outside the range $[0,1)$, and that $V(w)$ is negative in this range. Thus no physically allowed range for $u$ will arise from this subcase.\footnote{In the special case $\Phi=0$, we have $V(w\mathop{=}1)=0$ and so the single point $w=1$ {\it is\/} an allowed orbit. It corresponds to a particle that moves along either of the axes $\theta=0,\pi$. We do not consider such orbits in this paper.}  On the other hand, if $E^2>1$, the general shape of a quadratic function with negative leading coefficient allows for both roots $w_{1,2}$ to lie in the range $(0,1]$, so that $V(w)$ is non-negative somewhere in this range. Now, a necessary condition for both roots to be positive is\footnote{\label{footnote negative Q}This is, of course, not a sufficient condition. The reality of the square root in (\ref{w12}) will also impose a limit to how negative $Q$ can be. Nevertheless, checking the condition (\ref{Q<0 condition}) is sufficient for us to rule out spherical orbits with negative $Q$.}
\begin{align}
\label{Q<0 condition}
a^2(1-E^2)+Q+\Phi^2<0\,.
\end{align}
However, as we will see in Sec.~\ref{non-neg Q}, this condition will {\it not\/} be satisfied by the spherical orbits we find. It follows that $V(w)$ is negative everywhere in the range $[0,1]$. Hence, the case $Q<0$ will not occur in this paper.

\item{\underline{$Q=0$}:}\enskip If $E^2<1$, we have $w_1=0<1<w_2$. It follows that only equatorial orbits with $u=0$ are allowed. On the other hand, if $E^2>1$, non-equatorial orbits may be allowed, depending on the sign of $a^2(1-E^2)+\Phi^2$. If $a^2(1-E^2)+\Phi^2>0$, we have $w_2<w_1=0$ and again only equatorial orbits are allowed. If $a^2(1-E^2)+\Phi^2<0$, we have $w_2=0<w_1$ and non-equatorial orbits are allowed if $w_1<1$. However, since the condition $a^2(1-E^2)+\Phi^2<0$ is just a special case of (\ref{Q<0 condition}), this subcase will not occur for the spherical orbits we find.
\end{enumerate}

\section{Conditions for spherical orbits}
\label{section 3}

In order for a spherical orbit to exist at radius $r$, the conditions $R(r)=\frac{\dif R(r)}{\dif r}=0$ must hold at this radius. These two equations can be solved simultaneously, and the solutions take the most compact form when parameterised in terms of $r$ and $Q$. It turns out that there are four classes of solutions $(E_i,\Phi_i)$, which we label by $i=\text{a},\text{b},\text{c},\text{d}$. The first two are given by
\begin{subequations}
\label{solution12}
\begin{align}
\label{solution12a}
E_\text{a,b}&=\frac{r^3(r-2M)-a\big(aQ\mp\sqrt{\Upsilon}\big)}{r^2\sqrt{r^3(r-3M)-2a\big(aQ\mp\sqrt{\Upsilon}\big)}}\,,\\
\Phi_\text{a,b}&=-\frac{2Mar^3+(r^2+a^2)\big(aQ\mp\sqrt{\Upsilon}\big)}{r^2\sqrt{r^3(r-3M)-2a\big(aQ\mp\sqrt{\Upsilon}\big)}}\,,
\end{align}
\end{subequations}
where 
\begin{align}
\label{Upsilon}
\Upsilon\equiv Mr^5-Q(r-3M)r^3+a^2Q^2.
\end{align}
We will see below that these two classes of solutions are not actually separate solutions, but can be regarded as two different branches of the same solution. The third and fourth classes of solutions are related to the first two by
\begin{align}
\label{solution34}
(E_\text{c,d},\Phi_\text{c,d})=-(E_\text{a,b},\Phi_\text{a,b})\,.
\end{align}
We remark that these solutions have previously appeared in the literature \cite{Hughes:1999,Hughes:2001,Fayos:2007,Rana:2019}, albeit in different forms. In \cite{Hughes:1999}, they were parameterised in terms of $r$ and $\Phi$, while in \cite{Hughes:2001,Fayos:2007}, they were parameterised in terms of $r$ and $E$. In \cite{Rana:2019}, they were parameterised in terms of $r$ and $Q$, but in a form different from (\ref{solution12}).

As we will see in Sec.~\ref{reality2}, the first two classes of solutions have positive energy while the last two classes have negative energy. In the limit $Q=0$, the two positive-energy solutions (\ref{solution12}) reduce to those found in Eqs.~(2.12) and (2.13) of \cite{Bardeen:1972}, describing prograde and retrograde circular orbits in the equatorial plane, respectively.

On the other hand, in the limit of large $r$, we can read directly off from (\ref{solution12}) that
\begin{subequations}
\begin{align}
&E^2\approx\frac{(r-2M)^2}{r(r-3M)}\,,\\
&\Phi^2+Q\approx\frac{Mr^2}{r-3M}\,,
\end{align}
\end{subequations}
for both classes of solutions. These expressions are the squares of the energy and total angular momentum, respectively, of a circular orbit around a Schwarzschild black hole \cite{Wilkins:1972}. Thus for large $r$, the solutions (\ref{solution12}) describe (almost) circular orbits that are in general inclined with respect to the equatorial plane. 

From now on, we shall restrict our attention to the first two classes of solutions (\ref{solution12}), describing particles with positive energy. This is astrophysically the more relevant case, although recall that particles with negative energy can exist in the ergosphere of the Kerr black hole. If needed, the results we obtain can be extended to the negative-energy case by performing the sign changes in (\ref{solution34}).

We now work out the ranges of the parameters $r$ and $Q$ for which the solutions (\ref{solution12}) are valid. Note that there are two different square-root terms appearing in these solutions; imposing that they are real will lead to restrictions on the ranges of $r$ and $Q$. This will be done in Secs.~\ref{reality1} and \ref{reality2}. We also need to check if the condition (\ref{Q<0 condition}) holds when $Q$ is negative; this will be done in Sec.~\ref{non-neg Q} and we will find that $Q$ cannot be negative. The final ranges that we obtain are summarised in Sec.~\ref{summary ranges}.

\subsection{Reality of first square root}
\label{reality1}

To ensure that the smaller square root appearing in the solutions (\ref{solution12}) is real, we need to impose the condition
\begin{align}
\label{Upsilon>0}
\Upsilon\geq0\,.
\end{align}
Note from (\ref{Upsilon}) that $\Upsilon$ is quadratic in $Q$, and its two roots are
\begin{align}
\label{Q12}
Q_{1,2}\equiv\frac{r^2}{2a^2}\big(r(r-3M)\mp\sqrt{r\Xi}\big)\,,
\end{align}
where
\begin{align}
\label{Xi}
\Xi\equiv r^3-6Mr^2+9M^2r-4Ma^2.
\end{align}

Now, $\Xi$ is a cubic equation in $r$ that is familiar from the study of circular photon orbits in the equatorial plane around a Kerr black hole. The two largest roots of $\Xi$ are given by \cite{Bardeen:1972}
\begin{align}
\label{r12}
r_{1,2}\equiv 2M\bigg[1+\cos\bigg(\frac{2}{3}\arccos\bigg(\mp\frac{a}{M}
\bigg)\bigg)\bigg]\,,
\end{align}
and are the radii of the prograde and retrograde photon orbits, respectively. They lie in the ranges $M\leq r_1<3M<r_2\leq 4M$. The locations of these two photon orbits will play an important role in what follows, as they will demarcate the allowed radii of the spherical time-like orbits.

A useful fact to note is that $\Xi$ is negative in the range $r_1<r<r_2$. In particular, this means that real solutions for $Q_{1,2}$ only exist outside this range. It can be checked that $Q_1\leq Q_2<0$ when $r\leq r_1$, and $0<Q_1\leq Q_2$ when $r\geq r_2$.

Since $\Upsilon$ is a quadratic function of $Q$ with positive leading coefficient, we can straightforwardly deduce the ranges for which (\ref{Upsilon>0}) holds: When $r<r_1$ or $r>r_2$, the allowed ranges are $Q\leq Q_1$ and $Q\geq Q_2$. On the other hand, when $r_1\leq r\leq r_2$, there is no restriction on the range of $Q$.

\subsection{Reality of second square root}
\label{reality2}

For the solutions (\ref{solution12}) to be valid, the larger square root appearing in them also has to be real. Thus we need to impose the condition
\begin{align}
\label{Gamma>0}
\Gamma_\text{a,b}\equiv r^3(r-3M)-2a\big(aQ\mp\sqrt{\Upsilon}\big)\geq0\,.
\end{align}
Since the square root of $\Upsilon$ occurs in $\Gamma_\text{a,b}$, the parameter ranges arising from (\ref{Gamma>0}) should be a subset of those arising from (\ref{Upsilon>0}).

Furthermore, since the square roots of $\Gamma_\text{a,b}$ appear in the denominators of the solutions (\ref{solution12}), we shall impose the condition that they are non-zero. As the numerators of (\ref{solution12}) are in general non-zero, $\Gamma_\text{a,b}=0$ correspond to infinite-energy solutions. Such solutions describe null orbits, which are not the main focus of this paper. There is, however, one special case in which the numerators and denominators of (\ref{solution12}) are zero simultaneously. This is when $r\rightarrow r_1$ in the extremal case $a=M$, and it turns out that (\ref{solution12}) remain finite in this limit. Since these orbits appear to lie on the event horizon of the extremal Kerr black hole, they were called {\it horizon-skimming orbits\/} by Wilkins \cite{Wilkins:1972}. This special case will be considered in Appendix~\ref{horizon skimming}.

We first find the values of $r$ and $Q$ for which $\Gamma_\text{a,b}=0$. Since 
\begin{align}
\label{Gamma12}
\Gamma_\text{a}\Gamma_\text{b}=r^5\,\Xi\,,
\end{align}
we see that $\Gamma_\text{a}$ may vanish only if $r=r_1$ or $r_2$, and similarly for $\Gamma_\text{b}$. At either of these values of $r$, we have $Q_1=Q_2=\frac{r^3(r-3M)}{2a^2}$ and $\Upsilon=a^2(Q-Q_1)^2$. Thus $\Gamma_\text{a,b}$ reduce to
\begin{align}
\Gamma_\text{a,b}=2a^2\big(Q_1-Q\pm|Q_1-Q|\big)\,.
\end{align}
It follows that $\Gamma_\text{a}=0$ if $Q\geq Q_1$, and $\Gamma_\text{a}>0$ if $Q<Q_1$. On the other hand, $\Gamma_\text{b}=0$ if $Q\leq Q_1$, and $\Gamma_\text{b}<0$ if $Q>Q_1$. Since we want to impose $\Gamma_\text{a,b}>0$, it follows that the first class of solutions (a) is allowed only if $Q<Q_1$. The second class of solutions (b) is not allowed at all.

We now turn to the case when $r\neq r_{1,2}$. It follows from (\ref{Gamma12}) that for fixed $r\neq r_{1,2}$, the signs of $\Gamma_\text{a,b}$ do not change as $Q$ is varied. This is provided $\Gamma_\text{a,b}$ are continuous functions of $Q$, which is indeed the case if we assume the ranges of $Q$ obtained in Sec.~\ref{reality1}. It remains to check if $\Gamma_\text{a,b}$ are positive for a specific value of $Q$ in each of the allowed ranges. We have the following cases:

\begin{enumerate}[(i)]
\item{\underline{$r<r_1$ or $r>r_2$}:}\enskip The allowed ranges of $Q$ are $Q\leq Q_1$ and $Q\geq Q_2$. Note that $\Gamma_\text{a,b}=\sqrt{r^5\Xi}$ when $Q=Q_1$, and $\Gamma_\text{a,b}=-\sqrt{r^5\Xi}$ when $Q=Q_2$. It follows that $\Gamma_\text{a,b}>0$ if $Q\leq Q_1$, and $\Gamma_\text{a,b}<0$ if $Q\geq Q_2$. Thus, the range $Q\geq Q_2$ is ruled out for these cases.

\item{\underline{$r_1<r<r_2$}:}\enskip There is no restriction on the range of $Q$. Note that $\Gamma_\text{a,b}=\pm\sqrt{-r^5\Xi}$ when $Q=\frac{r^3(r-3M)}{2a^2}$. It follows that $\Gamma_\text{a}>0$ and $\Gamma_\text{b}<0$ for any value of $Q$. Thus, the second class of solutions (b) is not allowed in this case.
\end{enumerate}

Finally, we remark that the condition $\Gamma_\text{a,b}>0$ will imply that the numerator of (\ref{solution12a}) is positive. From (\ref{Gamma>0}), we have
\begin{align}
-a\big(aQ\mp\sqrt{\Upsilon}\big)>-\frac{1}{2}\,r^3(r-3M)\,,
\end{align}
so that
\begin{align}
r^3(r-2M)-a\big(aQ\mp\sqrt{\Upsilon}\big)>\frac{1}{2}\,r^3(r-M)>0\,.
\end{align}
Thus $(E_\text{a,b},\Phi_\text{a,b})$ are positive-energy solutions. It follows from (\ref{solution34}) that $(E_\text{c,d},\Phi_\text{c,d})$ are negative-energy solutions.

\subsection{Non-negativity of \texorpdfstring{$Q$}{Q}}
\label{non-neg Q}

The ranges of $Q$ that we have found from imposing the reality of the square roots in (\ref{solution12}) allow for it to be negative. Recall that there is physically a limit to how negative $Q$ can be for non-space-like geodesics \cite{Carter:1968}. Furthermore, as mentioned in Footnote \ref{footnote negative Q}, the reality of $w_{1,2}$ will also place a limit on how negative $Q$ can be. Notwithstanding these considerations, we will continue to assume the ranges of $Q$ obtained in the preceding section, and show that the condition (\ref{Q<0 condition}) is not satisfied by the entire class of solutions (\ref{solution12}). In particular, this will serve to rule out the case of negative $Q$ for spherical orbits with $E^2>1$. It will also rule out the possibility of non-equatorial orbits with $Q=0$.

Substituting (\ref{solution12}) into the left-hand side of (\ref{Q<0 condition}) gives 
\begin{align}
\label{Q<0 condition1}
a^2(1-E_\text{a,b}^2)+Q+\Phi_\text{a,b}^2=\frac{\Psi_\text{a,b}}{r^2\Gamma_\text{a,b}}\,,
\end{align}
where
\begin{align}
\Psi_\text{a,b}\equiv r^3(r-M)(Mr^3+a^2Q)-a^2\Upsilon+\big(Mr^3+a^2Q\mp2a\sqrt{\Upsilon}\big)^2.
\end{align}
The ranges of $Q$ that we have found in Sec.~\ref{reality2} ensure that $\Gamma_\text{a,b}$ are positive. It remains to show that $\Psi_\text{a,b}$ are also positive for these ranges. 

It can be checked, by explicit calculation, that $\Psi_\text{a}\Psi_\text{b}$ is quadratic in $Q$. For fixed $r$, $\Psi_\text{a}$ may vanish only at the roots of this quadratic; and similarly for $\Psi_\text{b}$. Now the discriminant of this quadratic is proportional to $-r^{15}\Delta^3$, which is negative; so both roots are actually complex in this case. Thus, for fixed $r$, $\Psi_\text{a,b}$ do not vanish for any real value of $Q$. Since $\Psi_\text{a,b}$ are continuous functions of $Q$ for the ranges we are interested in, it follows that their signs do not change as $Q$ is varied. In other words, if $\Psi_\text{a}$ is positive (negative) for a certain value of $r$ and $Q$, it remains positive (negative) as $Q$ is varied; and similarly for $\Psi_\text{b}$. In the following, we shall check that $\Psi_\text{a}$ and/or $\Psi_\text{b}$ are positive for a specific value of $Q$ in each of the allowed ranges obtained in Sec.~\ref{reality2}:

\begin{enumerate}[(i)]
\item{\underline{$r<r_1$ or $r>r_2$}:}\enskip The allowed range of $Q$ is $Q\leq Q_1$. Note that $\Psi_\text{a,b}=\frac{r^5}{2}[(\sqrt{r}(r-M)-\sqrt{\Xi})^2+2M\Delta]$ when $Q=Q_1$, which is manifestly positive. It follows that $\Psi_\text{a,b}>0$ if $Q\leq Q_1$.

\item{\underline{$r=r_1$ or $r_2$}:}\enskip The allowed range of $Q$ is $Q<Q_1$, although only the first class of solutions (a) is allowed. The above argument still holds by continuity, and it follows that $\Psi_\text{a}>0$ if $Q< Q_1$.

\item{\underline{$r_1<r<r_2$}:}\enskip There is no restriction on the range of $Q$, although only the first class of solutions (a) is allowed. Note that $\Psi_\text{a}=\frac{r^5}{2}[(\sqrt{r}(r-M)-\sqrt{-\Xi})^2+2M\Delta]$ when $Q=\frac{r^3(r-3M)}{2a^2}$, which is manifestly positive. It follows that $\Psi_\text{a}>0$ for any value of $Q$.
\end{enumerate}

Hence, we have shown that (\ref{Q<0 condition1}) is positive for all the ranges of $Q$ found in Sec.~\ref{reality2}. The condition (\ref{Q<0 condition}) does not hold, in particular, when $Q$ is negative. This will serve to rule out all $Q$ which lie in the negative range. The final allowed ranges of $Q$ are those obtained in Sec.~\ref{reality2} which lie in the non-negative range.

\subsection{Summary of parameter ranges}
\label{summary ranges}

We are now in a position to put the results of the preceding subsections together, and summarise the ranges of $r$ and $Q$ for which the solutions (\ref{solution12}) are valid:

\begin{enumerate}[(i)]
\item{\underline{$r<r_1$}:}\enskip The allowed range of $Q$ is $Q\leq Q_1$. But since $Q_1<0$, this case is ruled out altogether. 

\item{\underline{$r=r_1$}:}\enskip This case is similarly ruled out when $a<M$. However, when $a=M$, this case is allowed and will be considered separately in Appendix~\ref{horizon skimming}.

\item{\underline{$r_1<r<r_2$}:}\enskip Only the first class of solutions $(E_\text{a},\Phi_\text{a})$ is allowed. Initially $Q$ was allowed to take any value, but the results of Sec.~\ref{non-neg Q} cuts the allowed range to $0\leq Q<\infty$.

\item{\underline{$r=r_2$}:}\enskip Only the first class of solutions $(E_\text{a},\Phi_\text{a})$ is allowed, and the allowed range of $Q$ is $Q<Q_1$. Since $Q_1>0$ in this case, the final allowed range is $0\leq Q<Q_1$.

\item{\underline{$r>r_2$}:}\enskip Both classes of solutions $(E_\text{a,b},\Phi_\text{a,b})$ are allowed, and the allowed range of $Q$ is $Q\leq Q_1$ for either class. Again, since $Q_1>0$, the final allowed range is $0\leq Q<Q_1$ for either class. Note that at the maximum value $Q=Q_1$, we have $(E_\text{a},\Phi_\text{a})=(E_\text{b},\Phi_\text{b})$. In other words, the first and second classes of solutions meet at this point. This means that $(E_\text{a},\Phi_\text{a})$ and $(E_\text{b},\Phi_\text{b})$ can actually be regarded as two different branches of the {\it same\/} solution.
\end{enumerate}

The final results are also summarised in Table \ref{table ranges}.

\begin{table}[t]
\begin{center}
\begin{tabular}{cccc}
\hline
\hline
Range/value of $r$ & Solution class & Range of $Q$ & Remarks\\ 
\hline
$r<r_1$ & -- & -- & --\\
$r=r_1$ & a & $0\leq Q<\infty$ & ~Exists only when $a=M$\\
$r_1<r<r_2$ & a & $0\leq Q<\infty$ & --\\
$r=r_2$ & a & $0\leq Q<Q_1$ & --\\
$r>r_2$ & a,b & $0\leq Q\leq Q_1$ & --\\
\hline
\hline
\end{tabular}
\end{center}
\vskip-5pt
\caption{\label{table ranges}Parameter ranges for the solutions (\ref{solution12}). The case $r=r_1$ is only allowed when $a=M$, and will be considered in Appendix~\ref{horizon skimming}.}
\vskip5pt
\end{table}

\section{Properties of the spherical orbits}
\label{section 4}

In the preceding section, we found two classes or branches of solutions $(E_\text{a,b},\Phi_\text{a,b})$, given by (\ref{solution12}), describing spherical time-like orbits with positive energy. These solutions are parameterised in terms of $r$ and $Q$. Indeed, we have seen how the existence of these solutions depends on the value of $r$, as summarised in Table~\ref{table ranges}.

In this section, we will see how the properties of the spherical orbits depend on $Q$ (for fixed $r$). In particular, we will investigate the stability, energy and angular momentum of these orbits, in Secs.~\ref{stability}, \ref{energy} and \ref{angular momentum}, respectively. The results that we obtain are summarised in Sec.~\ref{summary properties}, and the parameter space is explicitly constructed in the extremal limit $a=M$.

\subsection{Stability}
\label{stability}

For the spherical orbits to be stable under perturbations in the radial direction, we require that $\frac{\dif^2 R(r)}{\dif r^2}<0$, where $R(r)$ is given by (\ref{R}). The threshold between stability and instability---the so-called {\it marginally stable\/} case---is then given by $\frac{\dif^2 R(r)}{\dif r^2}=0$. We now sketch how this equation can be solved.

Substituting the solutions $(E_\text{a,b},\Phi_\text{a,b})$ into $\frac{\dif^2 R(r)}{\dif r^2}$, we find that the resulting expressions can be written as $\frac{\Omega_\text{a,b}}{r^2\Gamma_\text{a,b}}$, where 
\begin{align}
\Omega_\text{a,b}\equiv-2Mr^5\Delta+8\big(Mr^3+a^2Q\mp a\sqrt{\Upsilon}\big)^2,
\end{align}
and $\Gamma_\text{a,b}$ are given by (\ref{Gamma>0}). It can be checked that $\Omega_\text{a}\Omega_\text{b}$ is quadratic in $Q$. For fixed $r$, $\Omega_\text{a}$ may vanish only at the roots of this quadratic, and similarly for $\Omega_\text{b}$.

It turns out that one of the roots of the quadratic is always negative, and so is not relevant for us. The other root is given by
\begin{align}
\label{Q_ms}
Q_\text{ms}\equiv-\frac{Mr^{5/2}\big[(\sqrt{\Delta}-2\sqrt{Mr})^2-4a^2\big]}{4a^2\big(r^{3/2}-M\sqrt{r}-\sqrt{M\Delta}\big)}\,,
\end{align}
and is non-negative when $r$ lies between the two real roots of the quartic equation $4r\Xi-3\Delta^2+12(Mr-a^2)^2=0$, explicitly given by Eq.~(2.21) of \cite{Bardeen:1972}.\footnote{\label{footnote Qs}These two roots are where $Q_\text{ms}$ vanishes, {\it except\/} in the extremal limit $a=M$. In this limit, $Q_\text{ms}$ remains positive at the smaller root, which coincides with the event horizon (c.f.~Fig.~\ref{param space 1} below).} Moreover, in the range $r\geq r_2$ where $Q_1$ is real, it satisfies $Q_\text{ms}\leq Q_1$; equality occurs when $r=r_\text{ms}^*$, where
\begin{align}
r_\text{ms}^*\equiv M\bigg[\frac{11}{4}+\frac{7}{2}\cos\bigg(\frac{1}{3}\arccos\bigg(\frac{143M^2+200a^2}{343M^2}\bigg)\bigg)\bigg]\,.
\end{align}

Explicit substitution shows that $\Omega_\text{a}$ vanishes when $Q=Q_\text{ms}$, only if $r\leq r_\text{ms}^*$. This marginally stable case occurs in the first branch of solutions. It can be checked that in this case, the spherical orbits in the first branch are stable when $0\leq Q<Q_\text{ms}$, and unstable when $Q>Q_\text{ms}$. If the first branch of solutions continues to the second branch at $Q=Q_1$, then the spherical orbits in the second branch are all unstable by continuity.

On the other hand, $\Omega_\text{b}$ vanishes when $Q=Q_\text{ms}$, only if $r\geq r_\text{ms}^*$. This marginally stable case occurs in the second branch of solutions. It can be checked that in this case, the spherical orbits in the second branch are unstable when $0\leq Q<Q_\text{ms}$, and stable when $Q_\text{ms}<Q\leq Q_1$. The spherical orbits in the first branch are then all stable by continuity.

\subsection{Energy}
\label{energy}

Recall from Sec.~\ref{r geodesic} that orbits are bound when $E^2<1$, and unbound when $E^2>1$. The marginally bound case occurs when $E^2=1$, and we would now like to solve this equation.

This equation can be solved in a way similar to the one used in the preceding subsection. If we calculate $E_\text{a,b}^2-1$, we find that the result can be written as $\frac{\Lambda_\text{a,b}}{r^4\Gamma_\text{a,b}}$, where 
\begin{align}
\Lambda_\text{a,b}\equiv-Mr^7+\big(2Mr^3+a^2Q\mp a\sqrt{\Upsilon}\big)^2,
\end{align}
and $\Gamma_\text{a,b}$ are given by (\ref{Gamma>0}). It can be checked that $\Lambda_\text{a}\Lambda_\text{b}$ is quadratic in $Q$. For fixed $r$, $\Lambda_\text{a}$ may vanish only at the roots of this quadratic, and similarly for $\Lambda_\text{b}$.

It turns out that one of the roots of the quadratic is always negative, and so is not relevant for us. The other root is given by
\begin{align}
\label{Q_mb}
Q_\text{mb}\equiv-\frac{Mr^2\big[r\big(\sqrt{r}-2\sqrt{M}\big)^2-a^2\big]}{a^2\big(\sqrt{r}-\sqrt{M}\big)^2}\,,
\end{align}
and is non-negative when $(\sqrt{M}+\sqrt{M-a})^2\leq r\leq(\sqrt{M}+\sqrt{M+a})^2$.\footnote{\label{footnote Qb}As in Footnote \ref{footnote Qs}, these two roots are where $Q_\text{mb}$ vanishes, except when $a=M$. In this limit, $Q_\text{mb}$ remains positive at the smaller root, which coincides with the event horizon (c.f.~Fig.~\ref{param space 1}).} Moreover, in the range $r\geq r_2$ where $Q_1$ is real, it satisfies $Q_\text{mb}\leq Q_1$; equality occurs when $r=r_\text{mb}^*$, where $r_\text{mb}^*$ is the largest real root of the quintic equation $r^2\Xi-M\Delta^2=0$.\footnote{This quintic equation cannot be solved in general. But when $a=M$, it reduces to a cubic equation, and the relevant root is given by
\begin{align*}
r_\text{mb}^*=\frac{M}{3}\bigg[5+2\sqrt{19}\cos\bigg(\frac{1}{3}\arccos\bigg(\frac{187}{722}\sqrt{19}\bigg)\bigg)\bigg]\simeq4.61M\,.
\end{align*}
\vspace{-7pt}
}

Explicit substitution shows that $E_\text{a}^2=1$ when $Q=Q_\text{mb}$, only if $r\leq r_\text{mb}^*$. This marginally bound case occurs in the first branch of solutions. It can be checked that in this case, the spherical orbits in the first branch are bound when $0\leq Q<Q_\text{mb}$, and unbound when $Q>Q_\text{mb}$. If the first branch of solutions continues to the second branch at $Q=Q_1$, then the spherical orbits in the second branch are all unbound by continuity.

On the other hand, $E_\text{b}^2=1$ when $Q=Q_\text{mb}$, only if $r\geq r_\text{mb}^*$. This marginally bound case occurs in the second branch of solutions. It can be checked that in this case, the spherical orbits in the second branch are unbound when $0\leq Q<Q_\text{mb}$, and bound when $Q_\text{mb}<Q\leq Q_1$. The spherical orbits in the first branch are then all bound by continuity.

\subsection{Angular momentum}
\label{angular momentum}

The angular momentum of the orbit $\Phi$ determines its motion about the $\phi$-axis. In particular, its sign will determine if the orbit is prograde or retrograde, as we shall see below. We now solve for the condition $\Phi=0$.

For the first branch of solutions, we find that $Q$ is given by
\begin{align}
\label{Q_0}
Q_0\equiv\frac{Mr^2\big(\Delta^2+4Mr^2(r-M)\big)}{(r^2+a^2)\big(r\Delta-M(r^2-a^2)\big)}\,.
\end{align}
It is non-negative, and thus relevant, when $r\geq r_0$, where
\begin{align}
r_0\equiv M+2\sqrt{\frac{3M^2-a^2}{3}}\cos\bigg(\frac{1}{3}\arccos\bigg(\frac{3M(M^2-a^2)}{3M^2-a^2}\sqrt{\frac{3}{3M^2-a^2}}\bigg)\bigg)\,.
\end{align}
Moreover, in the range $r\geq r_2$ where $Q_1$ is real, it satisfies $Q_0\leq Q_1$. It can be checked that $\Phi$ is positive when $Q<Q_0$, and is negative when $Q>Q_0$. On the other hand, no physically relevant solution to $\Phi=0$ exists in the second branch. If the first branch of solutions continues to the second branch at $Q=Q_1$, then $\Phi$ is negative everywhere in the second branch by continuity.

At first, one might expect that an orbit with positive/negative angular momentum $\Phi$ will have a $\phi$ coordinate that increases/decreases monotonically with proper time. However, this is not always the case for the spherical orbits we have found. On the other hand, we can consider the change $\Delta\phi$ over one complete oscillation in latitude of the orbit. The result can be found below in Eq.~(\ref{Delta_phi_bound}) for bound orbits, and Eq.~(\ref{Delta_phi_unbound}) for unbound orbits. It turns out that $\Delta\phi$ is positive when $\Phi>0$, and negative when $\Phi<0$. In this sense, orbits with $\Phi>0$ can be considered to be prograde, while those with $\Phi<0$ can be considered to be retrograde.

In the special case $\Phi=0$, the spherical orbit will reach---and indeed pass through---the poles $u=\pm1$. Such an orbit is known as a {\it polar orbit\/}. It can be obtained as a limit of either prograde orbits ($\Phi\rightarrow0^+$) or retrograde orbits ($\Phi\rightarrow0^-$).

\subsection{Summary of properties and parameter space}
\label{summary properties}

We are now in a position to put the results of the preceding subsections together, and summarise how the properties of the solution (\ref{solution12}) depend on $r$ and $Q$. This will be aided with explicit plots of the parameter space in the extremal limit $a=M$. Some comments about parameter space when $a<M$ will be made at the end of this section.

We have the following three main cases depending on the value of $r$:

\begin{enumerate}[(i)]
\item{\underline{$r_1\leq r<r_2$}:}\enskip Only the first branch of solutions $(E_\text{a},\Phi_\text{a})$ exists, and the allowed range of $Q$ is $0\leq Q<\infty$. The $(r,Q)$ parameter space for this branch of solutions in the limit $a=M$ is depicted in Fig.~\ref{param space 1}. The blue, red and green curves are the $Q=Q_\text{ms}$, $Q_\text{mb}$ and $Q_0$ curves, respectively. They divide the parameter space into stable/unstable, bound/unbound and prograde/retrograde regions, respectively. Thus, the dark gray area in Fig.~\ref{param space 1} corresponds to stable bound prograde orbits. The medium gray area corresponds to unstable bound orbits that may either be prograde or retrograde, depending on which side of the green curve it lies. The light gray area corresponds to unbound (and therefore unstable) orbits that may either be prograde or retrograde, again depending on which side of the green curve it lies.

Here, we shall focus on the case $r_1<r<r_2$, leaving the special case $r=r_1$ to Appendix~\ref{horizon skimming}. Now, the $Q=0$ line corresponds to stable prograde circular orbits in the equatorial plane. As $Q$ is increased (for fixed $r$), the energy of the orbit will increase. At the same time, the maximum latitude of the orbit $u_1$, given by (\ref{w12}), will also increase, so the orbit will no longer be equatorial. At $Q=Q_\text{ms}$, the orbit goes from being stable to being unstable. At the larger value $Q=Q_\text{mb}$, the orbit goes from being bound to being unbound. 

\begin{figure}[t]
  \begin{center}
    \includegraphics[scale=1]{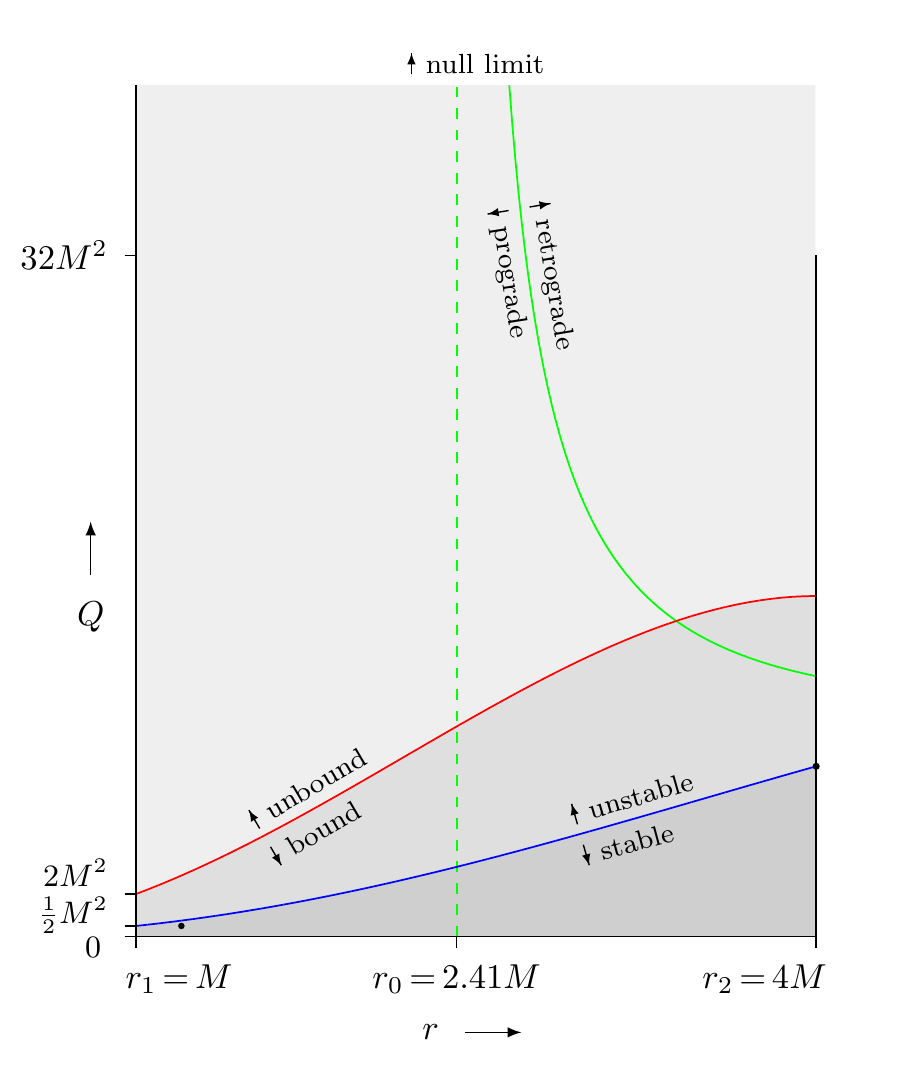}
  \end{center}
  \vskip-5pt
  \caption{\setstretch{1.2}The $(r,Q)$ parameter space for the first branch of solutions when $r_1\leq r\leq r_2$, in the limit $a=M$. The blue, red and green curves are the $Q=Q_\text{ms}$, $Q_\text{mb}$ and $Q_0$ curves, respectively. The green dashed line is the asymptote for the $Q=Q_0$ curve. The two black dots mark out orbits that will be illustrated in Sec.~\ref{section 6}.}
  \label{param space 1}
  \vskip5pt
\end{figure}

\begin{figure}
  \begin{center}
    \begin{subfigure}[b]{\textwidth}
      \centering
      \includegraphics[scale=1]{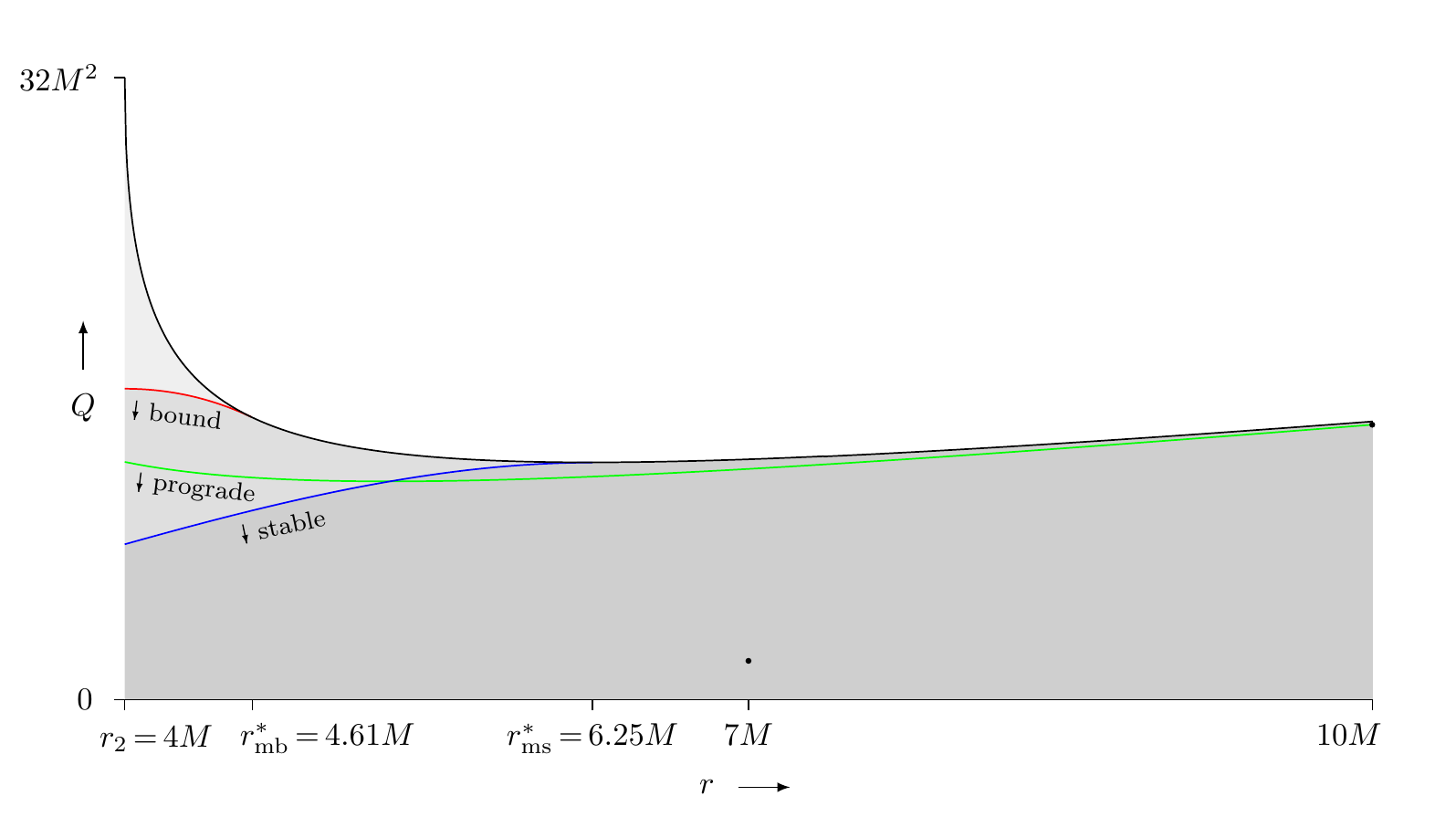}
      \caption{}
      \label{param space 2}
    \end{subfigure}
    \begin{subfigure}[b]{\textwidth}
      \centering
      \includegraphics[scale=1]{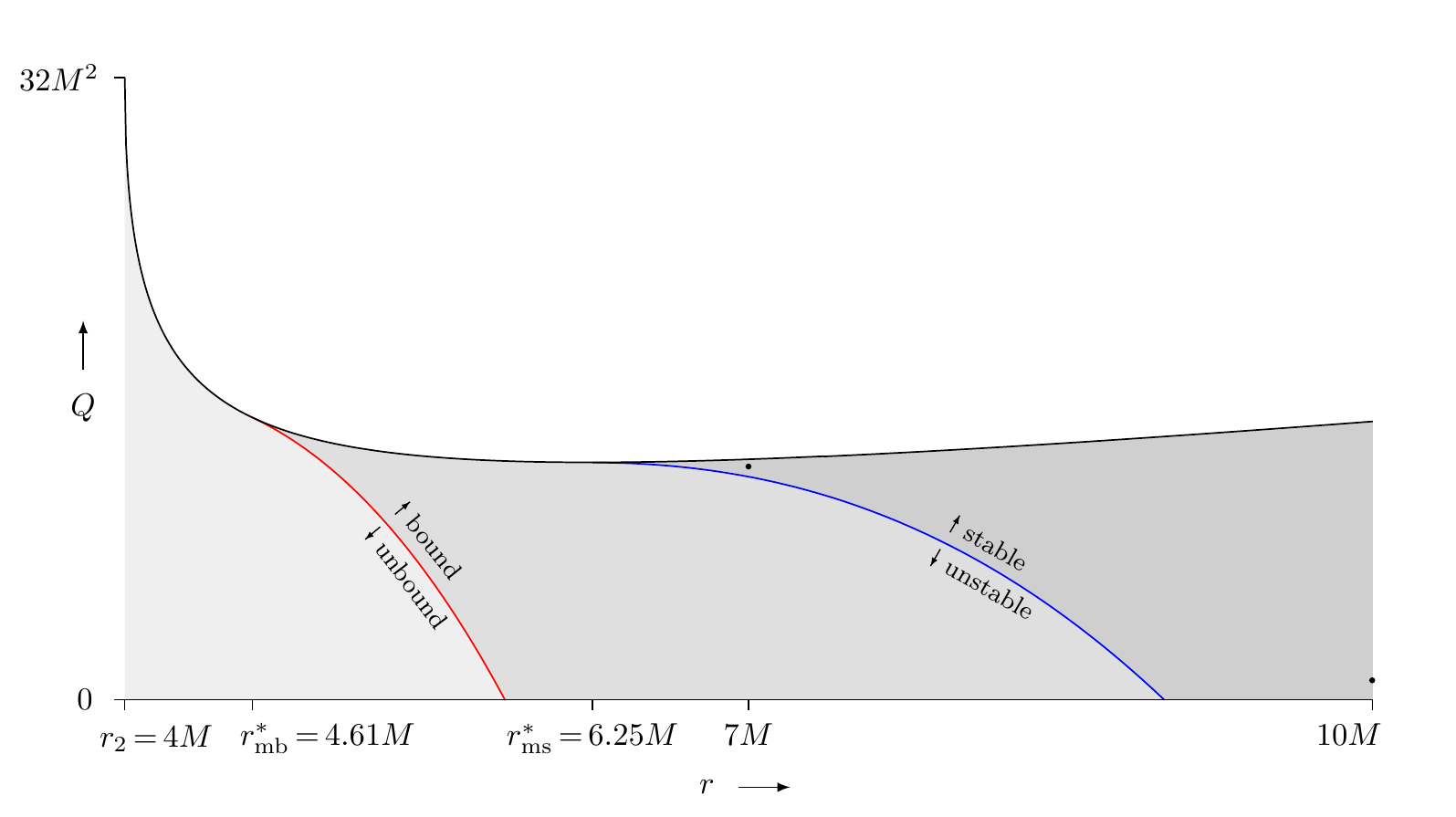}
      \caption{}
      \label{param space 3}
    \end{subfigure}
  \end{center}
  \vskip-5pt
  \caption{\setstretch{1.2}The $(r,Q)$ parameter space for (a) the first branch of solutions and (b) the second branch of solutions when $r>r_2$, in the limit $a=M$. The black curve is the $Q=Q_1$ curve, while the blue, red and green curves are as in Fig.~\ref{param space 1}. Due to space constraints, the unstable, unbound and retrograde regions of the parameter space are not labelled in (a). The orbits in (b) are all retrograde. The four black dots mark out orbits that will be illustrated in Sec.~\ref{section 6}.}
  \label{param space 23}
\end{figure}

The behaviour of the orbit's angular momentum will depend on whether $r\leq r_0$ or $r>r_0$. When $r\leq r_0$, the angular momentum will remain positive throughout. On the other hand, when $r>r_0$, the angular momentum will decrease to zero at $Q=Q_0$. At this point, the maximum latitude $u_1$ reaches the poles, and the orbit becomes polar. For $Q>Q_0$, the angular momentum becomes negative and the maximum latitude starts decreasing again; the orbits are now retrograde.

In the limit $Q\rightarrow\infty$, the energy $E$ will become infinite. However, the ratios $Q/E^2$ and $\Phi/E$ remain finite. This in fact corresponds to taking the null limit of our orbits, and they will reduce to the spherical photon orbits described in \cite{Teo:2003} (c.f.\ Appendix~\ref{appendix photon}). These spherical photon orbits will be prograde when $r<r_0$, and retrograde when $r>r_0$. It will be polar when $r=r_0$; indeed, substituting this value of $r$ into Eq.~(\ref{Phi photon}) below results in $\Phi=0$.

\item{\underline{$r=r_2$}:}\enskip As in the preceding case, only the first branch of solutions $(E_\text{a},\Phi_\text{a})$ exists, but now the allowed range of $Q$ is $0\leq Q<Q_1=32M^2$. This branch of solutions is represented by the black line on the right edge of the parameter space of Fig.~\ref{param space 1}. In the limit $Q\rightarrow32M^2$, the energy $E$ will become infinite. Again, this corresponds to taking the null limit of our orbits. Moreover, it can be checked that the maximum latitude of the orbit $u_1$ vanishes in this limit. Thus, this limit corresponds to none other than the retrograde circular photon orbit.

\item{\underline{$r>r_2$}:}\enskip Both branches of solutions $(E_\text{a,b},\Phi_\text{a,b})$ exist, and the allowed range of $Q$ for each branch is $0\leq Q\leq Q_1$. The $(r,Q)$ parameter space for each branch in the limit $a=M$ is depicted in Fig.~\ref{param space 23}. The black curve is the $Q=Q_1$ curve, where the two branches of solutions join up. The blue, red and green curves are as in Fig.~\ref{param space 1}, and they divide the parameter spaces into stable/unstable, bound/unbound and prograde/retrograde regions, respectively. The orbits in the second branch are all retrograde. Note that the red curve meets the black curve at $r_\text{mb}^*$, while the blue curve meets the black curve at $r_\text{ms}^*$; these are the points at which the red and blue curves cross over from one branch of solutions to the other.

Now, the $Q=0$ line in Fig.~\ref{param space 23}(\subref{param space 2}) corresponds to stable prograde circular orbits in the equatorial plane. As $Q$ is increased (for fixed $r$), the energy of the orbit will increase while its angular momentum will decrease. At the same time, the maximum latitude of the orbit $u_1$ will increase. At $Q=Q_0$, the angular momentum will decrease to zero, and the maximum latitude reaches the poles. For $Q>Q_0$, the angular momentum becomes negative and the maximum latitude starts decreasing again; the orbits are now retrograde. At larger value $Q=Q_1$, the orbit crosses over from the first branch of solutions to the second, and the value of $Q$ starts decreasing back to zero. The end-point $Q=0$ corresponds to a retrograde circular orbit in the equatorial plane.

In varying $Q$ as we did above, there could be a point at which $Q=Q_\text{ms}$; this could occur either in the first or second branch of solutions. At this point, the orbit goes from being stable to being unstable. There could also be a point at which $Q=Q_\text{mb}$, which could again occur either in the first or second branch of solutions. At this point, the orbit goes from being bound to being unbound. For $r\gtrsim5.83M$, the point $Q=Q_\text{mb}$ does not exist, and the orbits are all bound. For $r>9M$, the point $Q=Q_\text{ms}$ does not exist, and the orbits are all stable. When $r\rightarrow\infty$, we have $Q_0\rightarrow Q_1$, with $Q_1\rightarrow\infty$. As mentioned in Sec.~\ref{section 3}, circular Schwarzschild orbits are recovered in this limit.

\end{enumerate}

We remark that the full $(r,Q)$ parameter space for our solution can also be visualised as a two-dimensional surface embedded in a three-dimensional space spanned by, say, $r$, $Q$ and $\Phi$. This was done by Wilkins in Fig.~3 of \cite{Wilkins:1972}, for the extremal limit $a=M$, although he only considered the part of the parameter space corresponding to stable orbits. This figure makes it clear how the stable parts of Figs.~\ref{param space 1}, \ref{param space 23}(\subref{param space 2}) and \ref{param space 23}(\subref{param space 3}) join up smoothly together. It can be extended to include unstable bound and unbound orbits. However, note that when the unbound orbits of Fig.~\ref{param space 1} are included, the resulting two-dimensional surface will be infinitely extended along the $\Phi$ and $Q$ directions.

When $a<M$, the parameter space is qualitatively similar to that of the extremal limit $a=M$, but with two main differences. The first is that $r=r_1$ is no longer part of the parameter space. The second is that the ranges of $r$ for which $Q_\text{ms}$ and $Q_\text{mb}$ are non-negative will be smaller. The dependence of these ranges on $a$ can in fact be read off from Fig.~1 of \cite{Bardeen:1972}. For fixed $a<M$, the range of $r$ for which $Q_\text{ms}$ is non-negative lies between the two curves denoted by $r_\text{ms}$ in that figure. Similarly, the range of $r$ for which $Q_\text{mb}$ is non-negative lies between the two curves denoted by $r_\text{mb}$ in that figure. The two radii $r_1$ and $r_2$ themselves are denoted by $r_\text{ph}$ in that figure. It can be seen that, when $a=M/2$ for example, the region in which $Q_\text{ms}$ is non-negative lies entirely in the region $r>r_2$.

\section{Analytic solutions of the geodesic equations}
\label{section 5}

In this section, we provide analytic solutions of the geodesic equations for the spherical time-like orbits that we have obtained. Following Mino \cite{Mino:2003}, we first introduce a new parameter $\lambda$ along the geodesic, defined by $\frac{\dif\tau}{\dif\lambda}=\Sigma$. We then have
\begin{align}
\label{Mino}
\Sigma\frac{\dif x^\mu}{\dif\tau}=\frac{\dif x^\mu}{\dif\lambda}\,,
\end{align}
which can be used to simplify the left-hand side of each of the geodesic equations (\ref{eom}). The coordinates $(u,\phi,t)$ can then be solved in terms of $\lambda$ using elliptic integrals and Jacobi elliptic functions \cite{Fujita:2009}. Other recent works which solve the geodesic equations include \cite{Rana:2019,Kapec:2019,Gralla:2019,Compere:2020,Rana:2020}. We will use an approach similar to \cite{Kapec:2019,Gralla:2019,Compere:2020} to solve these equations, although there will be some differences in the way the final solutions are expressed. With these solutions in hand, the change in $\phi$ and $t$ for one complete oscillation in latitude can then be calculated. We will consider bound and unbound orbits separately.

\subsection{Bound orbits}

We begin by considering bound spherical orbits with $E^2<1$. Note that, in terms of the variable $w\equiv u^2$, and in terms of the new parameter $\lambda$, the geodesic equation for $u$, (\ref{u_eom}), can be written as
\begin{align}
\label{w_eom}
\frac{\dif w}{\dif\lambda}=\pm2Y(w)\,,
\end{align}
where
\begin{align}
Y(w)^2\equiv a^2(1-E^2)w(w_1-w)(w_2-w)\,.
\end{align}
Recall that $w_{1,2}$ is given by (\ref{w12}), with $0\leq w\leq w_1\leq1<w_2$. Without loss of generality, we choose the positive sign in (\ref{w_eom}) and assume the initial condition that $\lambda=0$ when $w=0$. Integrating (\ref{w_eom}) then gives \cite{G&R}
\begin{align}
\label{lambda_bound}
\lambda=\frac{1}{2}\int_0^w\frac{\dif w}{Y(w)}=\frac{1}{a\sqrt{(1-E^2)w_2}}\,F(\psi,k)\,,
\end{align}
where
\begin{subequations}
\begin{align}
\psi&\equiv\arcsin\sqrt{\frac{w}{w_1}}\,,\\
\label{k_bound}
k&\equiv\sqrt{\frac{w_1}{w_2}}\,,
\end{align}
\end{subequations}
and $F(\psi,k)$ is the incomplete elliptic integral of the first kind.\footnote{Our conventions for the elliptic integrals follow those of Gradshteyn and Ryzhik \cite{G&R}.} Thus we see that $\lambda$ monotonically increases with $w$, and reaches the value of $\lambda_0$ when $w=w_1$, where
\begin{align}
\lambda_0\equiv\frac{K(k)}{a\sqrt{(1-E^2)w_2}}\,,
\end{align}
and $K(k)$ is the complete elliptic integral of the first kind. At this parameter value, the orbit would have completed one-quarter of a complete oscillation in latitude.

To extend (\ref{lambda_bound}) past $\lambda=\lambda_0$, we first invert this equation using the Jacobi $\sn$ function \cite{A&S,Byrd}:
\begin{align}
\sin\psi=\sn\big(a\sqrt{(1-E^2)w_2}\,\lambda,k\big)\,.
\end{align}
In terms of the coordinate $u=\pm\sqrt{w}$, we can write this as\footnote{It is also possible to rewrite the right-hand side of this equation in terms of $u_1$ and $u_2$, which will lead to some simplifications in the present case. However, we will not do so as $u_2$ will be imaginary in the unbound case to be considered in Sec.~\ref{unbound}.}
\begin{align}
\label{u_bound}
u&=\sqrt{w_1}\,\sn\big(a\sqrt{(1-E^2)w_2}\,\lambda,k\big)\,.
\end{align}
$\psi$ itself can be written in terms of the Jacobi amplitude function as \cite{A&S,Byrd}
\begin{align}
\label{psi_bound}
\psi=\am\big(a\sqrt{(1-E^2)w_2}\,\lambda,k\big)\,.
\end{align}
Note that both (\ref{u_bound}) and (\ref{psi_bound}) are valid for any parameter value $\lambda$. A complete oscillation in latitude occurs within the range $0\leq\lambda<4\lambda_0$, and this is repeated with period $\Delta\lambda=4\lambda_0$.

The geodesic equations (\ref{phi_eom}) and (\ref{t_eom}) can similarly be integrated. We first use (\ref{Mino}) and (\ref{w_eom}) to rewrite them as
\begin{subequations}
\begin{align}
\frac{\dif\phi}{\dif w}&=\frac{1}{2Y(w)}\left[\frac{\Phi}{1-w}+\frac{a}{\Delta}(2MrE-a\Phi)\right],\\
\frac{\dif t}{\dif w}&=\frac{1}{2Y(w)}\left[-a^2E(1-w)+\frac{1}{\Delta}\big(E(r^2+a^2)^2-2Mra\Phi\big)\right].
\end{align}
\end{subequations}
Assuming the initial conditions that $\phi=t=0$ when $w=0$, these equations can then be integrated to give \cite{G&R}
\begin{subequations}
\label{phi_t_bound}
\begin{align}
\label{phi_bound}
\phi&=\frac{\Phi}{a\sqrt{(1-E^2)w_2}}\,\Pi\big(\psi,w_1,k\big)+\frac{a}{\Delta}(2MrE-a\Phi)\lambda\,,\\
t&=-\frac{aE}{\sqrt{(1-E^2)w_2}}\big[(1-w_2)F(\psi,k)+w_2E\big(\psi,k\big)\big]\cr
&\qquad+\frac{1}{\Delta}\big[E(r^2+a^2)^2-2Mra\Phi\big]\lambda\,,
\end{align}
\end{subequations}
where $E(\psi,k)$ and $\Pi(\psi,w_1,k)$ are the incomplete elliptic integrals of the second and third kind, respectively. $\psi$ and $k$ are again given by (\ref{psi_bound}) and (\ref{k_bound}), respectively. The solutions (\ref{phi_t_bound}) are valid for any parameter value $\lambda$, provided the elliptic integrals are understood to be extended outside their usual ranges. This is achieved by using the following symmetry and quasi-periodicity properties \cite{Byrd}:
\begin{subequations}
\begin{align}
F(-\psi,k)&=-F(\psi,k)\,,\\
E(-\psi,k)&=-E(\psi,k)\,,\\
\Pi(-\psi,w_1,k)&=-\Pi(\psi,w_1,k)\,,
\end{align}
\end{subequations}
and
\begin{subequations}
\begin{align}
F(\psi+\pi,k)&=F(\psi,k)+2K(k)\,,\\
E(\psi+\pi,k)&=E(\psi,k)+2E(k)\,,\\
\Pi(\psi+\pi,w_1,k)&=\Pi(\psi,w_1,k)+2\Pi(w_1,k)\,,
\end{align}
\end{subequations}
where $E(k)$ and $\Pi(w_1,k)$ are the complete elliptic integrals of the second and third kind, respectively. Note that our solutions are parameterised directly in terms of $\lambda$, instead of through $u$. Moreover, each coordinate is given by a single expression, rather than multiple expressions depending on the value of $\lambda$ or $u$. 

Note that $\phi$ as given by (\ref{phi_bound}) is allowed to take any value in the range $-\infty<\phi<\infty$, rather than being restricted to the range $0\leq\phi<2\pi$. This is useful for keeping track of how many revolutions an orbit makes around the black hole. Using (\ref{phi_t_bound}), we can calculate the change in $\phi$ and $t$ for one period $\Delta\lambda=4\lambda_0$, i.e., for one complete oscillation in latitude. We obtain
\begin{subequations}
\begin{align}
\label{Delta_phi_bound}
\Delta\phi&=\frac{4}{\sqrt{(1-E^2)w_2}}\bigg[\frac{\Phi}{a}\,\Pi(w_1,k)+\frac{2MrE-a\Phi}{\Delta}\,K(k)\bigg]\,,\\
\Delta t&=\frac{4}{\sqrt{(1-E^2)w_2}}\bigg[-aE\big((1-w_2)K(k)+w_2E(k)\big)\cr
&\qquad+\frac{E(r^2+a^2)^2-2Mra\Phi}{a\Delta}\,K(k)\bigg]\,.
\end{align}
\end{subequations}
The result for $\Delta\phi$ agrees with that obtained by Wilkins \cite{Wilkins:1972} in the extremal limit $a=M$. These results have also recently been obtained in \cite{Rana:2020}. Following the argument in \cite{Wilkins:1972}, it can be shown that $\Delta\phi$ is positive when $\Phi>0$, and negative when $\Phi<0$. Thus, the orbits are prograde when $\Phi>0$, and retrograde when $\Phi<0$. Moreover, when parameter values for various orbits are substituted in, one finds that $\Delta\phi>2\pi$ for prograde orbits, and $|\Delta\phi|<2\pi$ for retrograde orbits. When $\Phi=0$, there is a jump of exactly $4\pi$ in the graph of $\Delta\phi$, and the value of $\Delta\phi$ is taken to be the mid-point of this discontinuity \cite{Teo:2003}.

We remark that the solutions (\ref{u_bound}) and (\ref{phi_t_bound}) continue to be valid in the marginally bound case $E^2=1$. To see this, note that,
\begin{subequations}
\label{E=1_limit}
\begin{align}
w_1&=\frac{Q}{Q+\Phi^2}+O(1-E^2)\,,\\
w_2&=\frac{Q+\Phi^2}{a^2(1-E^2)}+O(1)\,.
\end{align}
\end{subequations}
In particular, the factor $\sqrt{(1-E^2)w_2}$ that appears in (\ref{u_bound}) and (\ref{phi_t_bound}) remains finite. The limiting forms of the elliptic integrals and Jacobi elliptic functions can also be readily obtained \cite{A&S,Byrd} (see also \cite{Kapec:2019,Compere:2020}).

\subsection{Unbound orbits}
\label{unbound}

We now turn to unbound spherical orbits with $E^2>1$. Recall that $w_2$ is now negative, so that we have $w_2<0\leq w\leq w_1\leq1$. The solutions for the bound case (\ref{u_bound}) and (\ref{phi_t_bound}) in fact continue to hold for the unbound case \cite{Kapec:2019,Compere:2020}. However, the elliptic integrals and Jacobi elliptic functions now have an imaginary elliptic modulus $k$ (or negative parameter $k^2$). 

Since the elliptic modulus is commonly taken to be real and in the range $0<k<1$, it might still be useful to present the solutions in a form that retains this property, which we shall do in this section. For example, it could facilitate comparison with other works (as we do in Appendix~\ref{appendix photon}). However, the resulting expressions will be different, and indeed somewhat longer, than those for the bound case.

With the new ranges of $E^2$ and $w_2$, the integral of the geodesic equation (\ref{w_eom}) now takes the form \cite{G&R}:
\begin{align}
\label{lambda_unbound}
\lambda=\frac{1}{2}\int_0^w\frac{\dif w}{Y(w)}=\frac{1}{a\sqrt{(E^2-1)(w_1-w_2)}}\,F(\psi,k)\,,
\end{align}
where
\begin{subequations}
\begin{align}
\psi&\equiv\arcsin\sqrt{\frac{w(w_1-w_2)}{w_1(w-w_2)}}\,,\\
\label{k_unbound}
k&\equiv\sqrt{\frac{w_1}{w_1-w_2}}\,.
\end{align}
\end{subequations}
Note that $k$ is real and lies in the range $0<k<1$, as desired. Inverting the equation (\ref{lambda_unbound}), we have
\begin{align}
\sin\psi=\sn\big(a\sqrt{(E^2-1)(w_1-w_2)}\,\lambda,k\big)\,,
\end{align}
which can then be used to give an expression for $u=\pm\sqrt{w}$ in terms of the Jacobi sd function \cite{A&S,Byrd}:
\begin{align}
\label{u_unbound}
u=\sqrt{-w_2}\,k\sd\big(a\sqrt{(E^2-1)(w_1-w_2)}\,\lambda,k\big)\,.
\end{align}
It follows that $u$ is a periodic function of $\lambda$, with period
\begin{align}
\Delta\lambda=\frac{4K(k)}{a\sqrt{(E^2-1)(w_1-w_2)}}\,.
\end{align}

The geodesic equations (\ref{phi_eom}) and (\ref{t_eom}) can similarly be integrated, to obtain \cite{Prudnikov}
\begin{subequations}
\label{phi_t_unbound}
\begin{align}
\phi&=\frac{\Phi}{a(1-w_2)\sqrt{(E^2-1)(w_1-w_2)}}\big[F(\psi,k)-w_2\Pi\big(\psi,k^2(1-w_2),k\big)\big]\cr
&\qquad+\frac{a}{\Delta}(2MrE-a\Phi)\lambda\,,\\
t&=-\frac{aE}{\sqrt{(E^2-1)(w_1-w_2)}}\big[(1-w_2)F(\psi,k)+w_2\Pi\big(\psi,k^2,k\big)\big]\cr
&\qquad+\frac{1}{\Delta}\big[E(r^2+a^2)^2-2Mra\Phi\big]\lambda\,,
\end{align}
\end{subequations}
where
\begin{align}
\psi=\am\big(a\sqrt{(E^2-1)(w_1-w_2)}\,\lambda,k\big)\,,
\end{align}
and $k$ is given by (\ref{k_unbound}). The change in $\phi$ for one period $\Delta\lambda$ is
\begin{align}
\label{Delta_phi_unbound}
\Delta\phi&=\frac{4}{\sqrt{(E^2-1)(w_1-w_2)}}\bigg[\frac{\Phi}{a(1-w_2)}\big(K(k)-w_2\Pi\big(k^2(1-w_2),k\big)\big)\cr
&\qquad+\frac{2MrE-a\Phi}{\Delta}\,K(k)\bigg]\cr
&=\frac{4}{\sqrt{(E^2-1)(w_1-w_2)}}\bigg[\frac{\Phi}{a(1-w_1)}\,\Pi\Big(-\frac{w_1}{1-w_1},k\Big)+\frac{2MrE-a\Phi}{\Delta}\,K(k)\bigg]\,,~~~~
\end{align}
where, in obtaining the second line, we have used Eq.~(17.7.17) of \cite{A&S}. The corresponding change in $t$ is
\begin{align}
\Delta t&=\frac{4}{\sqrt{(E^2-1)(w_1-w_2)}}\bigg[-aE\big((1-w_2)K(k)+w_2\Pi\big(k^2,k\big)\big)\cr
&\qquad+\frac{E(r^2+a^2)^2-2Mra\Phi}{a\Delta}\,K(k)\bigg]\,.
\end{align}
The behaviour of $\Delta\phi$ is similar to that in the bound case. In particular, the orbits are prograde when $\Phi>0$, and retrograde when $\Phi<0$.

\section{Example orbits}
\label{section 6}

Having obtained analytic solutions of the geodesic equations, we are now in a position to plot out several actual examples of spherical orbits around a Kerr black hole. They complement examples which have previously appeared in the literature (see, e.g., \cite{Goldstein:1974,Stoghianidis:1987,Hackmann:2010,Rana:2019,Rana:2020}). Here, we focus only on the case of stable or marginally stable orbits.

\begin{figure}
  \begin{center}
    \begin{subfigure}[b]{0.4\textwidth}
      \centering
      \includegraphics[scale=1]{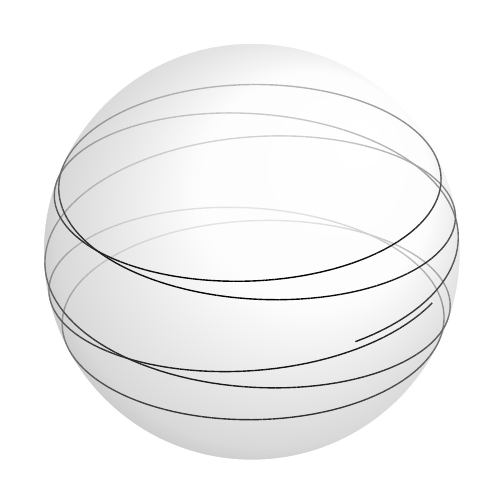}
      \caption{}
      \label{orbit (a)}
    \end{subfigure}~~~
    \begin{subfigure}[b]{0.4\textwidth}
      \centering
      \includegraphics[scale=1]{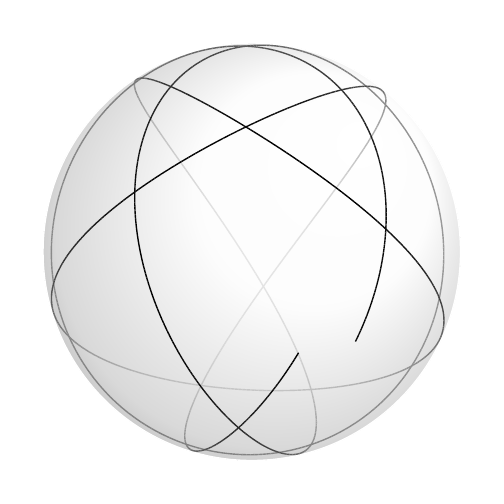}
      \caption{}
      \label{orbit (b)}
    \end{subfigure}
    \vskip1cm
    \begin{subfigure}[b]{0.4\textwidth}
      \centering
      \includegraphics[scale=1]{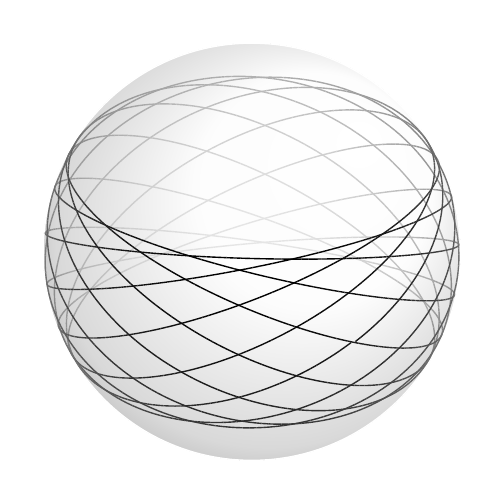}
      \caption{}
      \label{orbit (c)}
    \end{subfigure}~~~
    \begin{subfigure}[b]{0.4\textwidth}
      \centering
      \includegraphics[scale=1]{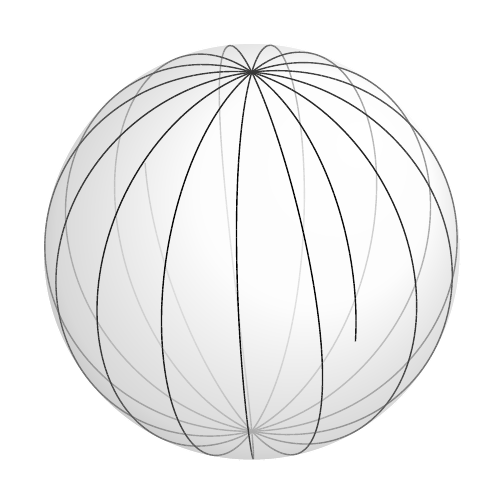}
      \caption{}
      \label{orbit (d)}
    \end{subfigure}
    \vskip1cm
    \begin{subfigure}[b]{0.4\textwidth}
      \centering
      \includegraphics[scale=1]{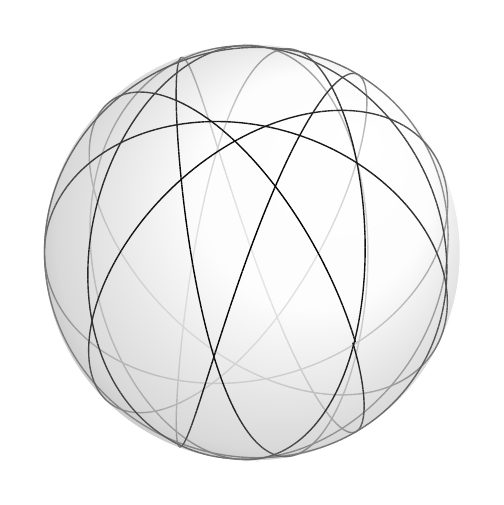}
      \caption{}
      \label{orbit (e)}
    \end{subfigure}~~~
    \begin{subfigure}[b]{0.4\textwidth}
      \centering
      \includegraphics[scale=1]{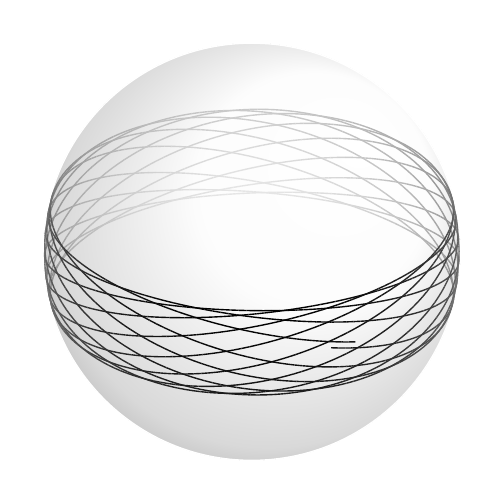}
      \caption{}
      \label{orbit (f)}
    \end{subfigure}
  \end{center}
  \vskip-5pt
  \caption{\setstretch{1.2}Six examples of spherical orbits, plotted on an imaginary sphere of fixed radius. The actual radii, as well as other properties of the orbits, are listed in Table~\ref{table}. Each orbit begins at the equator and heads northwards. The observer is located $30^\circ$ west of the starting point of the orbit, and $30^\circ$ north of the equator. The black hole itself rotates from west to east.}
  \label{example orbits}
\end{figure}

In Fig.~\ref{example orbits}, we have illustrated six example orbits around an extreme Kerr black hole. In each case, we plot the orbit on an imaginary sphere of fixed radius. (The actual radii, as well as other properties of the orbits, are listed in Table~\ref{table}.) Each orbit begins at the equator and heads northwards. The observer is located $30^\circ$ west of the starting point of the orbit, and $30^\circ$ north of the equator. The black hole itself rotates from west to east.

The first two orbits lie in the parameter space of Fig.~\ref{param space 1}. They are necessarily prograde orbits belonging to the first branch of solutions. The first is an orbit located close to the minimum radius $r_1$, which coincides with the radius of the event horizon in this case. For stability, the value of $Q$ for such an orbit has to be small. Note that such an orbit will in general have a large value of $\Delta\phi$, which means that it will make several revolutions around the black hole in one single oscillation in latitude. In this case, the orbit will make slightly more than five revolutions around the black hole in one latitudinal oscillation, as illustrated in Fig~\ref{example orbits}(\subref{orbit (a)}). 

The second orbit is located at the same radius as the retrograde circular photon orbit $r_2$. Its value of $Q$ is chosen to be $8M^2$, which makes it a marginally stable orbit. The energy of this orbit is, coincidentally, equal to its angular momentum (in units of $M$). It has a relatively high maximum latitude. Four latitudinal oscillations of this orbit are illustrated in Fig.~\ref{example orbits}(\subref{orbit (b)}).

\begin{table}
\begin{center}
\begin{tabular}{ccccccc}
\hline
\hline
Orbit & $r/M$ & $Q/M^2$ & $E$ & $\Phi/M$ & $u_1$ & $\Delta\phi$ \\
\hline
(a) & $1.2$ & $0.5$ & 0.71688 & $1.43571$ & 0.41062 & $31.94883$ \\
(b) & $4$ & $8$ & $0.91856$ & $0.91856$ & $0.95029$ & $7.77999$ \\
(c) & $7$ & $2$ & $0.93297$ & $2.62034$ & $0.47228$ & $6.85204$ \\
(d) & $10$ & $\frac{112900}{7979}$ & $0.95585$ & $0$ & $1$ & $0.39483$ \\
(e) & $7$ & $12$ & $0.95003$ & $-1.35045~~$ & $0.93127$ & $-5.58661~~$ \\
(f) & $10$ & $1$ & $0.96265$ & $-4.11659~~$ & $0.23560$ & $-5.84363~~$ \\
\hline
\hline
\end{tabular}
\end{center}
\vskip-5pt
\caption{\label{table}Properties of the spherical orbits illustrated in Fig.~\ref{example orbits}.}
\vskip5pt
\end{table}

The third and fourth orbits lie in the parameter space of Fig.~\ref{param space 23}(\subref{param space 2}), and belong to the first branch of solutions. The third orbit is a prograde one which is representative of the orbits lying in this region of the parameter space. Such orbits will have a value of $|\Delta\phi|$ that is only somewhat slightly greater than $2\pi$, which means that they will make slightly more than one revolution around the black hole in one latitudinal oscillation. Eleven latitudinal oscillations of this particular orbit are illustrated in Fig.~\ref{example orbits}(\subref{orbit (c)}). In this case, the orbit happens to end up very close to, but not at, the starting point. Continued plotting of this orbit will result in the filling up of the area between the latitudes $\pm u_1$.

The fourth orbit is an example of a polar orbit with zero angular momentum. Eight latitudinal oscillations of this particular orbit is illustrated in Fig.~\ref{example orbits}(\subref{orbit (d)}). Each latitudinal oscillation of the orbit looks to a certain extent like a great circle passing through the poles, but the ending point is slightly displaced from the starting point in the direction of the black hole's rotation. This is, of course, due to the dragging of inertial frames by the black hole.

The fifth and sixth orbits lie in the parameter space of Fig.~\ref{param space 23}(\subref{param space 3}), and belong to the second branch of solutions. They are necessarily retrograde orbits. Such orbits will have a value of $|\Delta\phi|$ that is somewhat slightly less than $2\pi$, which means that they will make slightly less than one revolution around the black hole in one latitudinal oscillation. The fifth orbit has a relatively high maximum latitude. Nine latitudinal oscillations of this orbit are illustrated in Fig.~\ref{example orbits}(\subref{orbit (e)}). In this case, the orbit happens to end up very close to, but not at, the starting point.

The sixth orbit has a value of $Q$ that is chosen to be small, and this results in a relativity low maximum latitude for the orbit. Fourteen latitudinal oscillations of this orbit are illustrated in Fig.~\ref{example orbits}(\subref{orbit (f)}). As $Q$ is decreased further, the maximum latitude will also decrease, and the orbit will approach an equatorial orbit.

Although we have only focussed on (marginally) stable spherical orbits, a random sampling of other orbits that are unstable or unbound with $Q\lesssim32M^2$ reveal similar features to those considered here. However, unbound orbits with $r_1<r<r_2$ and very large values of $Q$ are more similar to the spherical photon orbits considered in \cite{Teo:2003}. Also, orbits in the non-extremal case turn out to be qualitatively similar to those in the extremal case.

\section{Conclusion}

In this paper, we have made a systematic and thorough study of spherical time-like orbits around a Kerr black hole. Three main goals have been accomplished. Firstly, we have presented simplified forms of the energy $E$ and angular momentum $\Phi$ of the orbit, in terms of its radius $r$ and the value of Carter's constant $Q$. They are given by one of four solutions (\ref{solution12}) and (\ref{solution34}), although only the first two have positive energy. Focussing on the positive-energy solutions, we then worked out the ranges of $r$ and $Q$ for which these solutions are valid, as summarised in Table~\ref{table ranges}.

Secondly, we have studied how the properties of these orbits depend on these two parameters. For fixed $r$, we have found the value of $Q$, (\ref{Q_ms}), separating stable and unstable orbits. Similarly, we have found the value (\ref{Q_mb}) separating bound and unbound orbits, as well as the value (\ref{Q_0}) separating prograde and retrograde orbits.

Thirdly, we have provided analytic solutions of the geodesic equations for these orbits in terms of the Mino parameter using elliptic integrals and Jacobi elliptic functions. For bound orbits, they are given by (\ref{u_bound}) and (\ref{phi_t_bound}). These solutions are also formally valid for unbound orbits; however, the elliptic integrals and Jacobi elliptic functions now have an imaginary elliptic modulus. We have therefore presented alternative forms of the solutions, given by (\ref{u_unbound}) and (\ref{phi_t_unbound}), in which the elliptic modulus is real and lies between 0 and 1. 

The spherical time-like orbits considered in this paper have found important applications in astrophysics, especially in the study of gravitational waves from extreme mass ratio inspirals (EMRIs). In such a system, the orbit of the inspiralling body can be approximated by a sequence of Kerr geodesics, with the ``constants'' of motion evolving adiabatically due to radiation reaction. Now, it has been shown that spherical orbits remain spherical under radiation reaction \cite{Ryan:1995,Kennefick:1995}. This means that the evolution of such orbits will trace out a trajectory in the parameter space described in Sec.~\ref{section 4}. It would be interesting to understand the evolution and behaviour of these trajectories as a function of the starting point (initial set of orbital parameters), using phase space or other methods. The corresponding gravitational waveforms emitted can then be studied as in \cite{Hughes:1999,Hughes:2001}.

More generally, in the study of EMRIs, the location of the last stable orbit is of fundamental importance. Such orbits make up what is known as the {\it separatrix\/} in the parameter space of all orbits, since it separates the parameter space into those orbits that plunge into the black hole, and those that do not. These orbits are in fact homoclinic orbits---bound orbits which approach the same spherical orbit in the asymptotic past and future \cite{Levin:2008}. It turns out that homoclinic orbits are in one-to-one correspondence with bound but unstable spherical orbits. This allows the parameters of a homoclinic orbit to be related to those of the associated spherical orbit \cite{Rana:2019,Stein:2019}. Our simplified expressions for the parameters of spherical orbits (\ref{solution12}) would lead to correspondingly simplified expressions for the parameters of  homoclinic orbits. This opens up the possibility of an analytic characterisation of such orbits. They might also be helpful in speeding up numerical algorithms to locate the separatrix (although a very fast method not using homoclinic orbits was recently developed in \cite{Stein:2019}).

Finally, we mention that Rana and Mangalam \cite{Rana:2020} have very recently extended the relativistic precession model to spherical as well as non-equatorial eccentric orbits. They then applied it to the study of quasi-periodic oscillations (QPOs) in black hole X-ray binaries (BHXRBs). For two simultaneous QPO cases, they found spherical orbit solutions for two BHXRBs, namely M82 X-1 and XTEJ 1550-564. This is exciting evidence for the existence of spherical orbits around a Kerr black hole.

\section*{Acknowledgements}

I would like to acknowledge all the past students of the NUS Physics Department, who have contributed in one way or another to this project. I also wish to thank the reviewers for suggestions that have helped improve the presentation of the manuscript.

\appendix

\section{Horizon-skimming orbits}
\label{horizon skimming}

In \cite{Wilkins:1972}, Wilkins pointed out the existence of a class of so-called horizon-skimming orbits, which appear to lie on the event horizon of the extremal Kerr black hole with $a=M$. They arise by taking the $r\rightarrow r_1$ limit of the solution $(E_\text{a},\Phi_\text{a})$, and have the energy and angular momentum
\begin{align}
E=\frac{\sqrt{M^2+Q}}{\sqrt{3}M}\,,\qquad\Phi=2ME\,,
\end{align}
where $Q$ takes the range $0\leq Q<\infty$. These orbits are represented by the black line on the left edge of the parameter space of Fig.~\ref{param space 1}. The limit $Q\rightarrow\infty$ corresponds to taking the null limit of these orbits \cite{Teo:2003}.

The fact that the radii of these orbits coincide with that of the event horizon, is due to the well-known fact that the extremal Kerr black hole has an infinite throat in this region of the space-time \cite{Bardeen:1972}. Points along this throat share the same coordinate radius $r=M$, even though they might be (finitely or even infinitely) separated in space. Thus the horizon-skimming orbits remain above the event horizon; in fact, they also remain above the prograde circular photon orbit at $r_1$.

To resolve the throat region, we introduce a new parameter $\epsilon$ defined by
\begin{align}
\epsilon=\sqrt{1-\frac{a^2}{M^2}}\,.
\end{align}
The extremal limit is then taken as $\epsilon\rightarrow0$. We would like to understand the region of the parameter space near $r_1$ as this limit is taken. In particular, we focus on the marginally stable and marginally bound orbits in this region. Our results are consistent with those recently obtained in \cite{Compere:2020}.

Recall that marginally stable orbits are described by the curve $Q=Q_\text{ms}$. Substituting an expansion of the form $r=M+A\epsilon^p+\cdots$ into the right-hand side of this equation, and requiring that the lowest-order term is zeroth order in $\epsilon$, implies that $p=2/3$. The coefficient $A$ can then be determined in terms of $Q$, and we obtain \cite{Compere:2020}
\begin{align}
\label{r_ms}
r=M\Bigg[1+\bigg(\frac{M^2+Q}{M^2/2-Q}\bigg)^{1/3}\,\epsilon^{2/3}\Bigg]+O(\epsilon^{4/3})\,.
\end{align}
This parameterises the radii of these marginally stable orbits in terms of $Q$, which takes the range $0\leq Q<M^2/2$. The energy and angular momentum of these  orbits are given by
\begin{subequations}
\begin{align}
E&=\frac{\sqrt{M^2+Q}}{\sqrt{3}M}\Bigg[1+\bigg(\frac{\sqrt{2}(M^2-2Q)\epsilon}{M^2+Q}\bigg)^{2/3}\Bigg]+O(\epsilon^{4/3})\,,\\
\Phi&=2E+O(\epsilon^{4/3})\,.
\end{align}
\end{subequations}

On the other hand, recall that marginally bound orbits are described by the curve $Q=Q_\text{mb}$. Again, substituting an expansion of the form $r=M+A\epsilon^p+\cdots$ into the right-hand side of this equation, and requiring that the lowest-order term is zeroth order in $\epsilon$, implies that now $p=1$. The coefficient $A$ can then be determined in terms of $Q$, and we obtain \cite{Compere:2020}
\begin{align}
\label{r_mb}
r=M\bigg(1+\frac{2M\epsilon}{\sqrt{2M^2-Q}}\bigg)+O(\epsilon^2)\,.
\end{align}
This parameterises the radii of these marginally bound orbits in terms of $Q$, which takes the range $0\leq Q<2M^2$. The angular momentum of these orbits is given by \cite{Compere:2020}\footnote{Obtaining this result actually requires expanding $r$ in (\ref{r_mb}) to next order in $\epsilon$.}
\begin{align}
\Phi=2M+\sqrt{2M^2-Q}\,\epsilon+O(\epsilon^2)\,.
\end{align}

\begin{figure}[t]
  \begin{center}
    \includegraphics[scale=1]{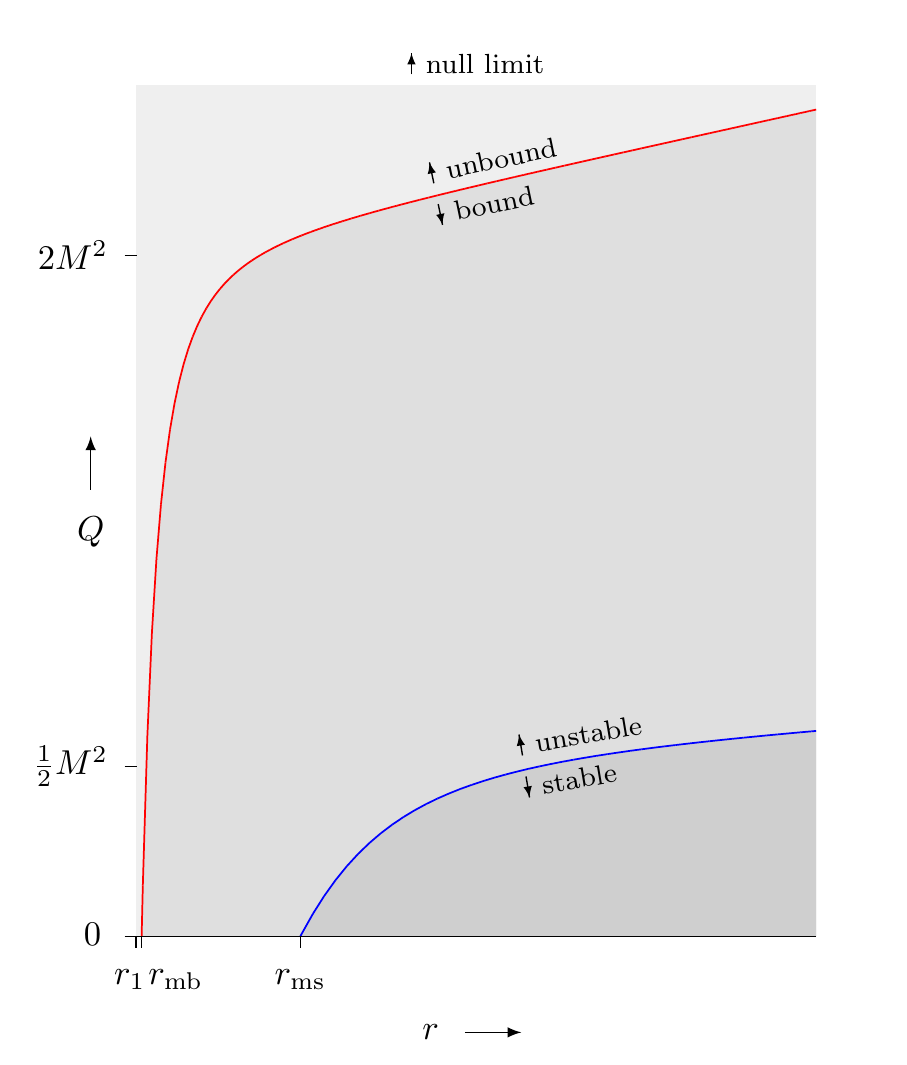}
  \end{center}
  \vskip-5pt
  \caption{\setstretch{1.2} The $(r,Q)$ parameter space near $r_1$ when $a=0.999995M$ (corresponding to $\epsilon\simeq0.003$). The blue and red curves are, as in Fig.~\ref{param space 1}, the $Q=Q_\text{ms}$ and $Q_\text{mb}$ curves, respectively.}
  \label{param space 4}
  \vskip5pt
\end{figure}

The $(r,Q)$ parameter space near $r_1$ is shown in Fig.~\ref{param space 4} for the case when $a=0.999995M$, corresponding to $\epsilon\simeq0.003$. The blue and red curves are, as in Fig.~\ref{param space 1}, the $Q=Q_\text{ms}$ and $Q_\text{mb}$ curves, respectively. They terminate on the $r$-axis at $r_\text{ms}$ and $r_\text{mb}$, respectively, if we borrow the notation of \cite{Bardeen:1972} in the equatorial limit. We would now like to understand what happens to this part of the parameter space, and in particular the two curves, when we take $\epsilon\rightarrow0$.

We begin with the red curve corresponding to marginally bound orbits. We have seen that the part of this curve for which $0\leq Q<2M^2$ is approximated by (\ref{r_mb}) when $\epsilon$ is small. In the limit $\epsilon\rightarrow0$, this part of the red curve gets ``flattened'' onto the black line on the left edge of Fig.~\ref{param space 1}, between $Q=0$ and $2M^2$. Thus, we see that although the red curve appears to terminate at the non-zero value $Q=2M^2$ in Fig.~\ref{param space 1}, it actually continues down to $Q=0$ along the black line. 

A similar situation happens for the blue curve corresponding to marginally stable orbits. The part of this curve for which $0\leq Q<M^2/2$ is approximated by (\ref{r_ms}) when $\epsilon$ is small. In the limit $\epsilon\rightarrow0$, this part of the curve gets ``flattened'' onto the same black line in Fig.~\ref{param space 1}, but now between $Q=0$ and $M^2/2$. Thus, the blue curve does not actually terminate at $Q=M^2/2$ in Fig.~\ref{param space 1}, but continues down to $Q=0$ along the black line.

It follows that the class of horizon-skimming orbits consists of at least a family of marginally bound orbits, and a family of marginally stable orbits, all sharing the same coordinate radius $r=M$. However, as mentioned above, these orbits are separated in space along the throat of the extremal Kerr black hole. In fact, it can be shown that the distance between $r_\text{mb}$ and $r_\text{ms}$ becomes infinite in the limit $\epsilon\rightarrow0$ \cite{Bardeen:1972}. The distance between $r_\text{ms}$ and the far regions of the space-time also becomes infinite in this limit. This is a manifestation of the fact that the throat is divided into distinct regions, as depicted in Fig.~2 of \cite{Bardeen:1972} (see also Fig.~1 of \cite{Kapec:2019}). The marginally bound orbits belong to one region (together with the photon orbit at $r_1$ and the event horizon itself), while the marginally stable orbits belong to another region. Other spherical orbits can also exist in these throat regions, and their locations relative to the marginally bound and marginally stable orbits are determined by the dependence of their radii on $\epsilon$. Geodesic motion in these throat regions have been the focus of recent attention in \cite{Kapec:2019,Compere:2020}.

\section{Spherical photon orbits}
\label{appendix photon}

In this appendix, we provide analytic solutions of the geodesic equations for the spherical photon orbits found in \cite{Teo:2003}. Recall that the null case corresponds to setting $\mu=0$ in (\ref{eom}) and (\ref{RU}). This case can also be recovered from the time-like case $\mu=1$, by taking the limit $E\rightarrow\infty$ of the solution $(E_\text{a},\Phi_\text{a})$ when $r_1<r<r_2$. As mentioned in Sec.~\ref{summary properties}, the ratios $\Phi/E$ and $Q/E^2$ remain finite in this limit. If we redefine $\Phi/E\rightarrow\Phi$ and $Q/E^2\rightarrow Q$, we arrive at the solution that was obtained in \cite{Teo:2003}:
\begin{subequations}
\begin{align}
\label{Phi photon}
\Phi&=-\frac{r\Delta-M(r^2-a^2)}{a(r-M)}\,,\\ 
Q&=-\frac{r^3\Xi}{a^2(r-M)^2}\,.
\end{align}
\end{subequations}
With these values of $\Phi$ and $Q$, $w_{1,2}$ can be calculated using
\begin{align}
w_{1,2}=\frac{1}{2a^2}\bigg[a^2-Q-\Phi^2\pm\sqrt{\big(a^2-Q-\Phi^2\big)^2+4a^2Q}\bigg]\,.
\end{align}
The coordinates $(u,\phi,t)$ of the geodesic can then be expressed in terms of the Mino parameter $\lambda$ as
\begin{subequations}
\begin{align}
u&=\sqrt{-w_2}\,k\sd\big(a\sqrt{w_1-w_2}\,\lambda,k\big)\,,\\
\phi&=\frac{\Phi}{a(1-w_2)\sqrt{w_1-w_2}}\big[F(\psi,k)-w_2\Pi\big(\psi,k^2(1-w_2),k\big)\big]\cr
&\qquad+\frac{a}{\Delta}(2Mr-a\Phi)\lambda\,,\\
t&=-\frac{a}{\sqrt{w_1-w_2}}\big[(1-w_2)F(\psi,k)+w_2\Pi\big(\psi,k^2,k\big)\big]\cr
&\qquad+\frac{1}{\Delta}\big[(r^2+a^2)^2-2Mra\Phi\big]\lambda\,,
\end{align}
\end{subequations}
where
\begin{align}
\psi=\am\big(a\sqrt{w_1-w_2}\,\lambda,k\big)\,,
\end{align}
and $k$ is given by (\ref{k_unbound}).

It follows that $u$ is a periodic function of $\lambda$, with period
\begin{align}
\Delta\lambda=\frac{4K(k)}{a\sqrt{w_1-w_2}}\,.
\end{align}
The change in $\phi$ and $t$ for one period $\Delta\lambda$ are
\begin{subequations}
\begin{align}
\Delta\phi&=\frac{4}{\sqrt{w_1-w_2}}\bigg[\frac{\Phi}{a(1-w_1)}\,\Pi\Big(-\frac{w_1}{1-w_1},k\Big)+\frac{2Mr-a\Phi}{\Delta}\,K(k)\bigg]\,,\\
\Delta t&=\frac{4}{\sqrt{w_1-w_2}}\bigg[-a\big((1-w_2)K(k)+w_2\Pi\big(k^2,k\big)\big)\cr
&\qquad+\frac{(r^2+a^2)^2-2Mra\Phi}{a\Delta}\,K(k)\bigg]\,.
\end{align}
\end{subequations}
We note that the result for $\Delta\phi$ agrees with that obtained in \cite{Teo:2003}.

\end{document}